\documentclass[iop,apj]{emulateapj}
\usepackage{amsmath,amssymb,amstext}
\usepackage{tikz}

\usepackage[breaklinks,colorlinks,citecolor=blue,linkcolor=magenta]{hyperref} 

\usepackage{mathtools}
\usepackage[slantedGreek,libertine]{newtxmath}
\usepackage{aas_macros}
\usepackage{float}
\usepackage{array}
\usepackage{enumitem}
\usepackage{booktabs}
\usepackage[normalem]{ulem}
\newcolumntype{L}[1]{>{\raggedright\let\newline\\\arraybackslash\hspace{0pt}}m{#1}}
\newcolumntype{C}[1]{>{\centering\let\newline\\\arraybackslash\hspace{0pt}}m{#1}}
\newcolumntype{R}[1]{>{\raggedleft\let\newline\\\arraybackslash\hspace{0pt}}m{#1}}

\shorttitle{Clump size measurements}
\shortauthors{Ali et al.}

\newcommand{\Msun}{\,\mathrm{M}_\odot}

\begin{document}

\title{Robust cross-correlation based measurement of clump sizes in galaxies}

\author{Kamran Ali$^1$, Danail Obreschkow$^1$, David B.~Fisher$^2$, Karl Glazebrook$^2$, Ivana Damjanov$^{5}$, Roberto G.~Abraham$^{3,4}$, Robert Bassett$^2$}

\affil{$^1$International Centre for Radio Astronomy Research (ICRAR), M468, University of Western Australia, 35 Stirling Hwy, Crawley, WA 6009, Australia \\
$^2$Centre for Astrophysics and Supercomputing, Swinburne University of Technology, P.O. Box 218, Hawthorn, VIC 3122, Australia\\
$^3$Department of Astronomy and Astrophysics, University of Toronto, 50 St. George St., Toronto, ON M5S 3H8, Canada\\
$^4$Dunlap Institute, University of Toronto, 50 St. George St., Toronto, ON M5S 3H8, Canada\\
$^5$Harvard-Smithsonian Center for Astrophysics, 60 Garden St., Cambridge, MA 02138, USA}

\begin{abstract}
Stars form in molecular complexes that are visible as giant clouds ($\sim 10^{5-6} \Msun$) in nearby galaxies and as giant clumps ($\sim 10^{8-9} \Msun$) in galaxies at redshifts $z\approx1$$-$$3$. Theoretical inferences on the origin and evolution of these complexes often require robust measurements of their characteristic size, which is hard to measure at limited resolution and often ill-defined due to overlap and quasi-fractal substructure. We show that maximum and luminosity-weighted sizes of clumps seen in star formation maps (e.g.\ H$\alpha$) can be recovered statistically using the two-point correlation function (2PCF), if an approximate stellar surface density map is taken as the normalizing random field. After clarifying the link between Gaussian clumps and the 2PCF analytically, we design a method for measuring the diameters of Gaussian clumps with realistic quasi-fractal substructure. This method is tested using mock images of clumpy disk galaxies at different spatial resolutions and perturbed by Gaussian white noise. We find that the 2PCF can recover the input clump scale at $\sim20\%$ accuracy, as long as this scale is larger than the spatial resolution. We apply this method to the local spiral galaxy NGC 5194, as well as to three clumpy turbulent galaxies from the DYNAMO-HST sample. In both cases, our statistical H$\alpha$-clump size measurements agree with previous measurements and with the estimated Jeans lengths. However, the new measurements are free from subjective choices when fitting individual clumps.
\end{abstract}

\maketitle

\section{Introduction}

Most stars formed roughly 8-12 Gyr ago, in galaxies now seen at redshifts $z\approx1$$-$$3$ (\citealp{Hopkins2006ApJ...651..142H}). Typical star-forming galaxies at these redshifts show a much more irregular and clumpy structure than local main-sequence galaxies (\citealp{Elmegreen2004ApJ...604L..21E}, \citealp{Elmegreen2006ApJ...650..644E}). The dominant mechanism (\citealp{Shibuya2016ApJ...821...72S}) leading to such structures is the in-situ structure formation due to violent instabilities within the galactic disk (\citealp{Dekel2009ApJ...703..785D}, \citealp{Bournaud2014ApJ...780...57B}). However, a significant portion of the observed clumpiness could very well be a result of mergers, both major and minor (\citealp{Ribeiro2016arXiv161105869R}). Hence, differentiating between the two formation scenarios (see details in \citealp{Fisher2017ApJ...839L...5F}), using scaling relations between clump sizes and other properties (\citealp{Wisnioski2012}, \citealp{Livermore2012MNRAS.427..688L}), could answer fundamental questions on the cosmic history of star formation.

In practice, measuring cloud/clump sizes is challenging. The star forming regions generally exhibit quasi-fractal (approximately fractal defined on a finite domain) substructures with scales extending from the Jeans length down to individual newborn stars. Hence, the observed distribution of sizes depends on the spatial resolution, and the ``characteristic'' size requires a proper definition, since the simple average size monotonically decreases with increasing resolution (\citealp{Fisher2017MNRAS.464..491F}). Additionally, observing noise and point-spread functions affect different scales in different ways. Current methods (see \citealp{Glazebrook2013PASA...30...56G}) hardly address these challenges: clumps are normally identified visually or as peaks above a preset noise threshold (typically $\sim$$3\sigma$, \citealp{Bassett2017MNRAS.467..239B}, \citealp{Jones2010MNRAS.404.1247J}). Clump sizes are then derived by fitting or associating the peaks with analytical shapes (e.g.\ Gaussians, ellipses, circles, isophotes). Such methods inherently depend on the observing noise, spatial resolution \citep{Pleuss2000A&A...361..913P} and subjective choices, making the comparison of different samples and different authors (e.g.~different redshifts and observations versus simulations) cumbersome.

Motivated by the success of the two-point correlation function (2PCF, \citealp{Peebles1980lssu.book.....P}) in measuring scales in cosmology, this paper develops and tests a two-point statistics for measuring the characteristic sizes of star-forming complexes -- more precisely we aim to measure the characteristic scale of the ``primary'' clumps, irrespective of their fragmented substructure. The 2PCF has already been fruitful in measuring the geometric distribution of stars (\citealp{Sanchez2010ApJ...720..541S}), correlating star formation tracers and ages (\citealp{Scheepmaker2009A&A...494...81S}),identifying truncation scales of galactic disks (\citealp{Combes2012A&A...539A..67C}) and theoretical modeling (\citealp{Hopkins2012MNRAS.423.2016H}). Unlike in these previous works, measuring clump sizes brings the extra complication that other galactic structures (e.g.\ exponential disk, spiral arms, bars) can substantially interfere with clump sizes, especially for the large clumps at high $z$. We approach this problem by normalizing the 2PCF of the star formation map (e.g.\ H$\alpha$ map) relative to the stellar surface density (e.g.\ continuum map), mimicking the way cosmological 2PCFs are corrected for complex survey selection functions.

This paper first reviews the functional form of the 2PCF of clumps with a Gaussian profile. It then describes a method of extracting a primary clump size (section \ref{sec:background}) and tests this method using mock data (section \ref{sec:statistics}). Section \ref{sec:M51} details the application of this method to the galaxy NGC 5194 as a test bed to quantify the robustness of our method. In section \ref{sec:DYNAMO}, we apply this method to DYNAMO-\textit{HST} (\textit{Hubble Space Telescope}) galaxy sample and compare our results with previously computed clump sizes. 

\section{Background and Idea} \label{sec:background}

The key idea is to use the spatial 2PCF of a galaxy image to characterize the clump sizes. This image is now treated as a 2D density field $\delta(\pmb{r})$, where $\pmb{r}\in\mathbb{R}^2$ is the position vector within the image. The 2PCF of this field is defined as
\begin{equation}
	\xi(r)=\overline{\delta(\pmb{t})\delta(\pmb{t}+\pmb{r})},
\end{equation}
where the overline denotes the average over all possible translations $\pmb{t}\in \mathbb{R}^2$ and rotations, such that $r\equiv|\pmb{r}|$. Although individual galaxies only have a single density field $\delta(\pmb{r})$, it is mathematically convenient to consider ensembles $\{\delta(\pmb{r})\}$ of statistically identical fields. In this case the 2PCF becomes
\begin{equation}
	\left\langle\xi(r)\right\rangle=\overline{\left\langle\delta(\pmb{t})\delta(\pmb{t}+\pmb{r})\right\rangle},
\end{equation}
where $\langle\rangle$ is the ensemble average. It is often easier to evaluate 2PCF in Fourier space, where random translations $\pmb{t}$ reduce to random phase factors that disappear in the ensemble averages. The Fourier transform (FT) of the 2PCF is called the power spectrum and defined as
\begin{equation} \label{eq:defps}
	\left\langle p(k)\right\rangle = \overline{\left\langle P(\pmb{k})\right\rangle} = \overline{\left\langle \delta(\pmb{k})\delta^{\dagger}(\pmb{k})\right\rangle},
\end{equation}
where $p(k)$ and $P(\pmb{k})$ are the isotropic (rotationally averaged) and non isotropic power spectra and $\delta(\pmb{k})$ is the FT of $\delta(\pmb{r})$ (defined in Appendix \ref{app:expectation}). It follows from Eq.~(\ref{eq:defps}) (see Appendix \ref{app:expectation}) is that a summed field $\delta(\pmb{r})=\sum_{\rm l=1}^N\delta_l(\pmb{r})$ of $N$ density fields has the power spectrum
\begin{equation} \label{eq:ps}
\begin{split}
\left\langle P(\pmb{k})\right\rangle & = \left\langle\sum_{\rm l=1}^{N}P_l(\pmb{k})\right\rangle + \left\langle\sum_{\rm l=1}^{N}\sum_{\rm \substack{m=1 \\ m\neq l}}^{N}\delta_l(\pmb{k})\delta^{\dagger}_m(\pmb{k})\right\rangle,\\
\end{split}
\end{equation}
where the second term is the cross-correlation between fields $\delta_l$ and $\delta_m$. Hence, the power spectrum of a sum of uncorrelated fields is simply the sum of the individual power spectra -- a property that we will exploit hereafter.

\subsection{Gaussian clump and weighted Two point function} \label{gauss2wpf}

Let us first consider a simple model of a single clump given by a 2D symmetric Gaussian, $\delta(\pmb{r})\propto e^{-(\pmb{r}-\pmb{\mu})^2/(2\sigma^2)}$, with standard deviation $\sigma$ and a random center $\pmb{\mu}$. This field has the useful property that the power spectrum and 2PCF are also Gaussians, centered at the origin and with standard deviations $(2\sigma^2)^{-1/2}$ and $(2\sigma^2)^{1/2}$, respectively (see Appendix \ref{app:renormal}). According to Eq.~(\ref{eq:ps}), a density field composed of \emph{many} 2D Gaussians with identical $\sigma$, but randomized positions, then has a Gaussian 2PCF 
\begin{equation}
\langle\xi_{\rm \sigma}(r)\rangle \propto e^{-\frac{r^2}{2\left(\sqrt{2}\sigma\right)^2}},
\end{equation}
with standard deviation $\sqrt{2}\sigma$. In other words, the size $\sigma$ of randomly positioned 2D Gaussian clumps can be recovered, exactly, by fitting a 1D Gaussian profile of standard deviation $\sqrt{2}\sigma$ to $\langle\xi(r)\rangle$.

If we deal with only a single density field $\delta(\pmb{r})$ (not a statistical ensemble) composed of multiple 2D Gaussian clumps, the particular 2PCF, $\xi(r)$, can deviate from a pure Gaussian due to non vanishing random cross-correlations between the individual clumps. In this case, fitting a Gaussian to $\xi(r)$ is not necessarily the best way to recover $\sigma$. A more suitable statistical measure is the weighted 2PCF ($\mathit{w}$2PF), $r^\gamma\xi(r)$, with positive exponent $\gamma>0$.
~This function exhibits the convenient property that its maximum position is proportional to the clump size, $r_{\rm peak} = \sqrt{2\gamma}\sigma$. In particular, if $\gamma=1/2$, the Gaussian clump size can be measured as $\sigma=r_{\rm peak}$. As we will show in Section \ref{sec:statistics}, this way of determining $\sigma$ is more robust against random cross-correlations between individual clumps. Moreover we will show that this method also produces good results if the density field is more complex, e.g.\ composed of differently sized clumps and clumps with realistic substructure.

\subsection{Numerical Estimator} \label{sec:LS}

The 2PCF of a galaxy image $\delta(\pmb{r})$ not only depends on the clump structure, but is also affected by other features, such as central bars, spiral arms and the overall decline in the surface density with radius. It is important to remove these additional effects in order to extract the characteristic clump size. This challenge is analogous to measuring the cosmological 2PCF of galaxies with complex selection functions. In cosmology, this problem is usually solved by constructing a random density field $R$ with the same selection function as the galaxy density field $D$, but uniformly distributed galaxies (no clustering).  The 2PCF is then estimated by using the expression of  \citealp{LS1993} (henceforth the LS-estimator):
\begin{equation} \label{eq:LS}
\hat{\xi}_{\rm LS}(r)= \frac{DD(r)-2DR(r)+RR(r)}{RR(r)},
\end{equation}
where the functions $DD$, $DR$ and $RR$ are defined as
\begin{equation} \label{eq:normalise}
\begin{split}
XY(r)\equiv\frac{1}{\sum X\sum Y} \sum_{\rm |\pmb{r_1-r_2}|\in(r\pm\Delta r/2)} X(\pmb{r_1})Y(\pmb{r_2}).
\end{split}
\end{equation}
The parameter $\Delta r$ is the bin width of the regularly distributed scale lengths $r$. Equation (\ref{eq:LS}) effectively removes the spurious 2PCF from the selection function, already present in the $R$-field, and retains only the 2PCF in the $D$-field not yet present in the R-field (for details refer to \citealp{LS1993}).

In the present case, the features of interest are the length scales of star-forming clumps. However, the maps of a star formation tracer (e.g.~an emission line image of ionized or molecular gas, UV continuum image, radio synchrotron image, etc.) also show other structures, such as  the exponential disk profile, spiral arms, central bars, etc. These other structures appear as a contamination of the 2PCF, if we are only interested in the clump scales. A possible solution, employed by \citealp{Zhang2001ApJ...561..727Z}, consists of smoothing out the flux map by a Gaussian filter (FWHM $\approx 3$ kpc) and subtracting the 2PCF of this map from the 2PCF of the star formation tracer. This tends to give a flat small-scale correction and a monotonically decreasing large-scale correction to the stellar 2PCF. The result is primarily due to choosing a smoothing scale much larger than the correlation scale of interest. We take a different approach, however, by choosing the map of the older stellar population (e.g.~an optical or near IR continuum image) as the $R$-field and computing the full LS-estimator. In this way, the $R$-field contains the galaxy's structure other than the star-forming clumps, and hence removes all this other structure from the 2PCF when using Eq.~(\ref{eq:LS}). In practice, the map of the older stellar population ($R$-field) might contain some clump structure, too, which means that a part of the clump signal is removed as well. This primarily reduces the amplitude of the 2PCF, so that the impact on the measurement of clump size is expected to be small (following our method below).

In summary, our method to determine the characteristic size of star-forming clumps works as follows:
\begin{enumerate}
\item prepare the images $D$ (star formation tracer) and $R$ (old stellar population),
\item compute $\hat{\xi}_{\rm LS}(r)$ via Eq.~(\ref{eq:LS}),
\item fit an offset $c$, such that $\tilde{\xi}(r)\equiv\hat{\xi}_{\rm LS}(r)-c$ vanishes at large $r$,
\item find the maximum position $r_{\rm peak}$ of $\sqrt{r}\tilde{\xi}(r)$.
\end{enumerate}

The third step is required because $\hat{\xi}_{\rm LS}(r)$ doesn't vanish at large $r$ in the case of a finite number of clumps (see Appendix \ref{app:renormal}). We perform the fitting of $r_{\rm peak}$ (step 4) at sub-$\Delta r$ accuracy, by fitting a parabola to the maximum three points of the $\mathit{w}$2PF, $\sqrt{r}\tilde{\xi}(r)$. As discussed in Section \ref{sec:background}, $r_{\rm peak}$ is identical to the clump size $\sigma$ in the simplistic case of equally sized Gaussian clumps at uncorrelated random positions. In reality, clumps come in different sizes and they have correlated (fractal-like) substructure. The meaning of $r_{\rm peak}$ in these cases will be explored numerically in the next Section.

\section{Clump size measurements in Mock Data} \label{sec:statistics}

In this Section, we connect the estimator $r_{\rm peak}$ to the size of clumps using mock images for the $D$-fields. All these images consist of $N_{\rm clumps}=10$ randomly placed clumps with periodic boundary conditions (see top row of Table \ref{tbl:distribution}). For the geometry of the clumps we consider three different models that are increasingly realistic. The first and simplest clump model consists of 2D Gaussians with identical size $\sigma$ (Section \ref{subsec:identical}) -- this case was already mentioned in the previous Section. While this clump model is far from realistic, it provides some analytical insight and helps to gauge the accuracy of our method. The second clump model still assumes that each clump is a 2D Gaussian function, but their sizes $\sigma$ are drawn from a \emph{power-law} distribution (Section \ref{subsec:PL}). This distribution is frequently used to relate the size of H$\alpha$ regions to their luminosity and number (e.g.~\citealp{Kennicutt1980ApJ...241..573K}, \citealp{Zurita2001Ap&SS.276..491Z}). Finally, motivated by the observed quasi-fractal structure of star-forming clouds (\citealp{Scheepmaker2009A&A...494...81S}, \citealp{Sanchez2008ApJS..178....1S}), we consider a more complex clump model, where each clump has quasi-fractal \emph{substructure} (Section \ref{subsec:SS}). These three clump models have different parameters, namely the clump size, power-law exponent and substructure properties.

Our mock density fields are generated on a grid of $100\times100$ pixels. This resolution and the number of clumps roughly mimic the images of the clumpy galaxies analyzed in Section \ref{sec:DYNAMO}. The corresponding $R$-fields are taken to be uniform. For each of the three clump models, Table \ref{tbl:distribution} shows one realization of the $D$-field, with the corresponding LS-estimator ($\hat{\xi}_{\rm LS}(r)$) and the $\mathit{w}$2PF ($\sqrt{r}\tilde{\xi}(r)$).

In all cases, the expectations of the 2PCFs are monotonically decreasing with $r$, but for a single random realization with a finite number of clumps, this function typically shows slight oscillatory behaviour. In the presence of such random oscillations we find $r_{\rm peak}$ (i.e.~the value of $r$ that maximizes $\sqrt{r}\tilde{\xi}(r)$) to be a more robust estimator of the clump sizes than some functional fits to the raw 2PCF. In the following Sections (\ref{subsec:identical}--\ref{subsec:SS}) we illustrate and test this idea by generating ensembles of random $D$-fields (similar to those shown in Table \ref{tbl:distribution}(a)) for various parameter settings of the three clump models. The ensemble-averaged values of $r_{\rm peak}$ with their standard deviations are shown in Table \ref{tbl:distribution}(d).

Finally, we look at systematics introduced into the $\mathit{w}$2PF by a Gaussian PSF and Gaussian white noise. 

\subsection{Gaussian clumps with equal sizes} \label{subsec:identical}

In the first model (Table \ref{tbl:distribution}, left column), clumps of identical sizes $\sigma_0$ and fluxes are distributed randomly in the 2D plane. In calculating the 2PCF, the total flux is renormalized as in Eq.~(\ref{eq:normalise}). In the limit of infinitely many clumps, where the cross-correlation term in Eq.~(\ref{eq:ps}) vanishes, the expected 2PCF is simply identical to that of a single Gaussian clump i.e.~the 2PCF is simply $\langle\xi_{\rm \sigma_0}(r)\rangle$. This conclusion holds for a finite number of clumps, up to an additive constant in $\langle\xi_{\rm \sigma_0}(r)\rangle$ coming from the nonvanishing cross-correlation term. This additive constant is automatically removed when measuring the 2PCF of a real clump image (step 3 in section \ref{sec:LS}).

Following Section \ref{gauss2wpf}, the expectation of the estimator $r_{\rm peak}$ is exactly equal to $\sigma_0$. The numerical analysis shows that for a single realization, $r_{\rm peak}$ matches the value of $\sigma_0$ within a standard deviation of $\lesssim 20\%$ and no measurable systematic error (Table \ref{tbl:distribution}(d), left).

\subsection{Gaussian clumps with Power-law Size Distribution} \label{subsec:PL}

As in the previous model, we here consider randomly positioned Gaussian clumps, but their sizes $\sigma$ are now drawn from a power-law distribution
\begin{equation}\label{eq:pwl}
	\phi(\sigma) \propto \sigma^{\beta},~\text{if}~0 \leq \sigma \leq \sigma_{\rm max}
\end{equation}
where $\sigma_{\rm max}$ represents the size of the largest clumps ($0<\sigma_{\rm max}<\infty$) and $\beta$ is the power-law exponent. 
The luminosities of the clumps are assumed to scale as a power law
\begin{equation}
L(\sigma)\propto\sigma^\omega,
\end{equation}
with power-law index $\omega$. Observations and theory typically find values of $\beta\approx-4..-3$ (\citealp{Oey2003AJ....126.2317O}, \citealp{Guszejnov2016MNRAS.459....9G}) and $\omega\approx2.7..3$ (\citealp{Wisnioski2012}, \citealp{Stromgren1939ApJ....89..526S}).

The 2PCF (for infinitely many clumps) is calculated as
\begin{equation} \label{eq:PL}
\begin{split}
	\left\langle\xi(r)\right\rangle &\propto \int_0^{\sigma_{\rm max}}\phi(\sigma)~L(\sigma)^2~\xi_{\rm \sigma}(r)~d\sigma\\
	& \propto \frac{1+\alpha}{2^\alpha\sigma^{1+\alpha}_{\rm max}} r^{\alpha-1}\ \Gamma\left(\frac{1-\alpha}{2},\frac{r^2}{4\sigma^2_{\rm max}}\right),
	\end{split}
\end{equation}
where $\alpha\equiv\beta+2\omega$ and $\Gamma$ is the upper incomplete Gamma function. This 2PCF is a monotonically decreasing function of $r$, which asymptotes to a power law near the origin.

Eq.~(\ref{eq:PL}) shows that the 2PCF depends only on $\alpha$, with no additional dependence on $\beta$ and $\omega$. With the aim of relating this 2PCF to a characteristic clump size, defined in some explicit way, it therefore makes sense to identify an average clump size that depends only on $\alpha$, not on any other combination of $\beta$ and $\omega$. We find that the average size of the clumps weighted by $L^q$ 
\begin{equation}
	\begin{split} 
	\bar{\sigma}_q &= \frac{\int_{\rm 0}^{\sigma_{\rm max}} \phi(\sigma)~L(\sigma)^q~\sigma~d\sigma}{\int_{\rm 0}^{\sigma_{\rm max}}\phi(\sigma)~L(\sigma)^q~d\sigma},\\
	&= \frac{1+(\beta+q\omega)}{2+(\beta+q\omega)}\sigma_{\rm max} 
	\end{split}
\end{equation} 
depends on $\sigma_{\rm max}$ and $\beta+q\omega$. Hence the $L^2$-weighted clump size depends only on $\alpha$ and $\sigma_{\rm max}$,
\begin{equation} \label{eq:PLsigma}
	\bar{\sigma}_{\rm 2}= \frac{1+\alpha}{2+\alpha}\sigma_{\rm max}.
\end{equation}
Note that Eqs.~(\ref{eq:PL}) and (\ref{eq:PLsigma}) only apply if $\alpha>-1$, which is always the case observationally ($\alpha\approx1.4..3$, according to the values above). Given a particular realization of this 2PCF, how can we extract a clump size? It turns out that finding the maximum $r_{\rm peak}$ of the $\mathit{w}$2PF is again a fruitful approach: the 2PCF of Eq.~(\ref{eq:PL}) is unbound for $r\rightarrow0$, but the $\mathit{w}$2PF is finite and has a nonzero maximum $r_{\rm peak}$, as long as $\alpha>0.5$ (satisfied by the observations quoted above). Numerically (see Table \ref{tbl:distribution}, (d) middle), we find that for $\alpha \geq 3$, the value of $r_{\rm peak}$ closely matches $\bar{\sigma}$ within a standard deviation within $20\%$ and negligible systematic error. For smaller values of $\alpha$, $r_{\rm peak}$ tends to underestimate $\bar{\sigma}$ and the standard deviation becomes higher (up to $\approx 40\%$).

\subsection{Clumps with Quasi-fractal Substructure} \label{subsec:SS}

The main difference between this model and previous ones is the spatial correlation between substructures. The construction of the mock $D$-field starts with the generation of randomly positioned, equally sized, Gaussian structures which we refer to as the \textit{primary} clumps. We then generate $N_{\rm sub}$ Gaussian substructures within each primary clump at random positions drawn from the 2D Gaussian profile of the primary clump. These structures are called first-generation clumps. This process is repeated recursively within each sub-clump to generate $N_{\rm sub}^2$ second-generation clumps, $N_{\rm sub}^3$ third-generation clumps and so on. The relative flux in substructure is set by the user-parameter $f \in (0,1)$, such that in every clump a fraction $f$ of its total flux is contained in substructure, while a faction $1-f$ remains in the Gaussian of that clump. The second user-parameter is the relative clump size $s\in (0,1)$ between consecutive clump generations. The $L^q$-weighted average clump size of this model takes the expression of a geometric series, which solves to
\begin{equation}
	\bar{\sigma}_{\rm q}= \frac{1-f^q}{1-sf^q}\sigma_{\rm max} \ \forall \ q>0.
\end{equation}
where $\sigma_{\rm max}$ is the size of the primary clump. While generating mocks for this model we ensure at least $95\%$ of total flux is generated in every realization i.e.~we require more generations for larger $f$ values.

Visually, this model mimics the clump structure often seen in disk galaxies. This, of course, is not a coincidence, because fragmentation of Jeans instability follows a similar rule where a collapsing structure produces more unstable regions. Our simple model is designed to mirror this recursive production of collapsing regions.

Although the density field of this model has a simple expression, it is difficult to write down the analytic form of the 2PCF. This is due to the presence of the nonvanishing correlation terms. Furthermore, fitting a sum of sequential Gaussians to the raw 2PCF is not a good approach. How can we extract a meaningful clump size for such a quasi fractal distribution? We find again that computing the $\mathit{w}$2PF and finding its peak location gives good results. Table \ref{tbl:distribution}, (d) right, shows that the numerically extracted value of $r_{\rm peak}$ tends to measure the size of the primary clump, $\sigma_{\rm max}$, rather than the $L^2$-weighted average size of the distribution with standard deviation within $20\%$. This is a desirable result because, observationally, using conventional methods on a resolved data set results in fitting smaller clumps and ignoring their overall distribution scale, while $r_{\rm peak}$, on the other hand, should still retain information of this larger correlation length scale.

\newcommand{\topvalue}{0.1cm}
\newcommand{\widthvalue}{0.25}
\newcommand{\widthvalueimage}{0.2}
\newcommand{\imagepaste}[1]{\hspace*{0.25cm} \includegraphics[trim=0 0 0 0cm, clip=true,width=0.26\textwidth]{#1} } 
\newcommand{\graphpaste}[1]{ \includegraphics[trim=0 0 0.2 -0.2cm, clip=true,width=0.26\textwidth]{#1} }

\begin{figure*}
\centering
     \begin{center}
     \begin{tabular}{p{2cm} c c c }
     \toprule
       Distribution & Identical Clumps & Power-law & substructure \\  \bottomrule
      \toprule\\
      \vspace{-2.5cm}  
       \begin{tabular}{p{0.15cm} p{1.5cm}}
       (a) & Data Field
       \end{tabular}
	&
       \imagepaste{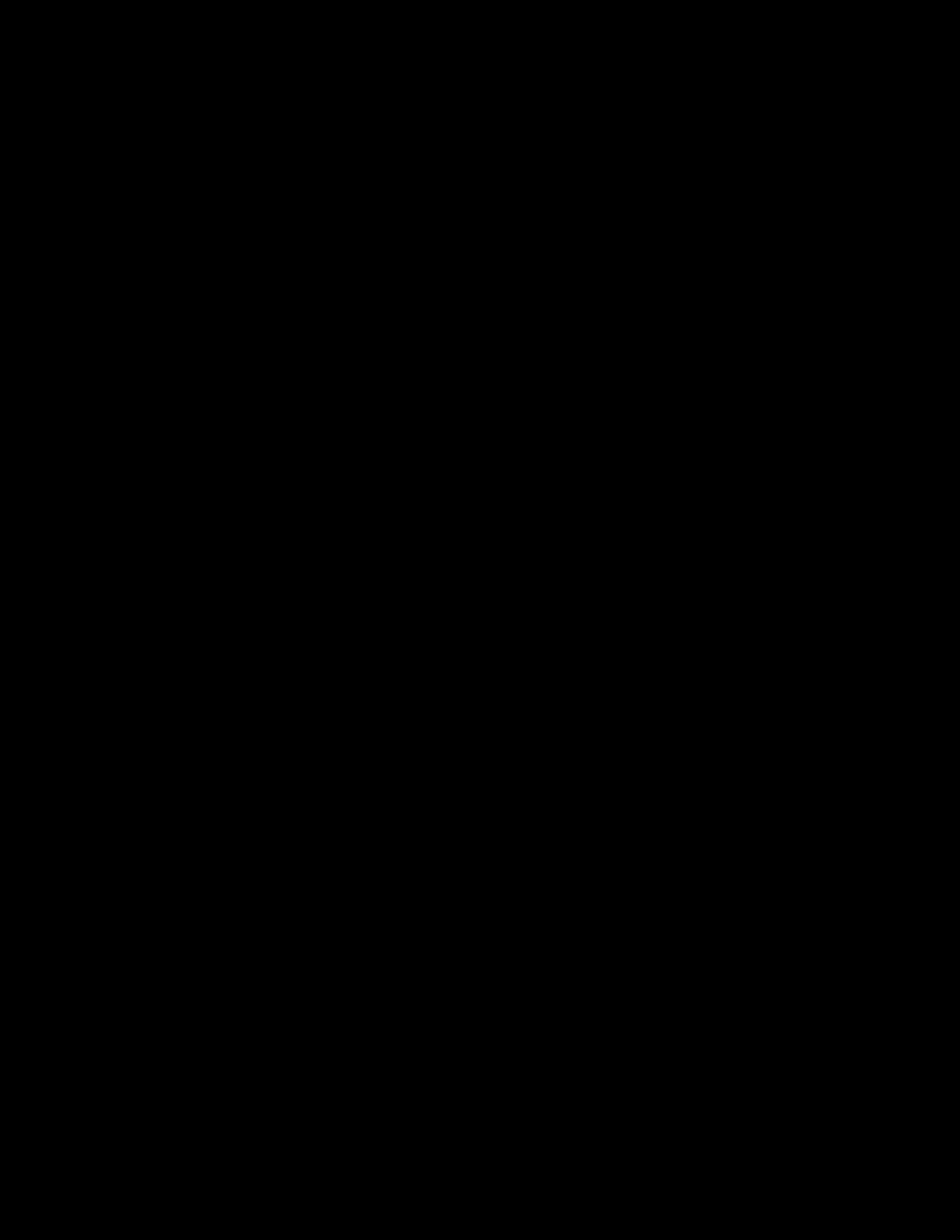}
      & 
       \imagepaste{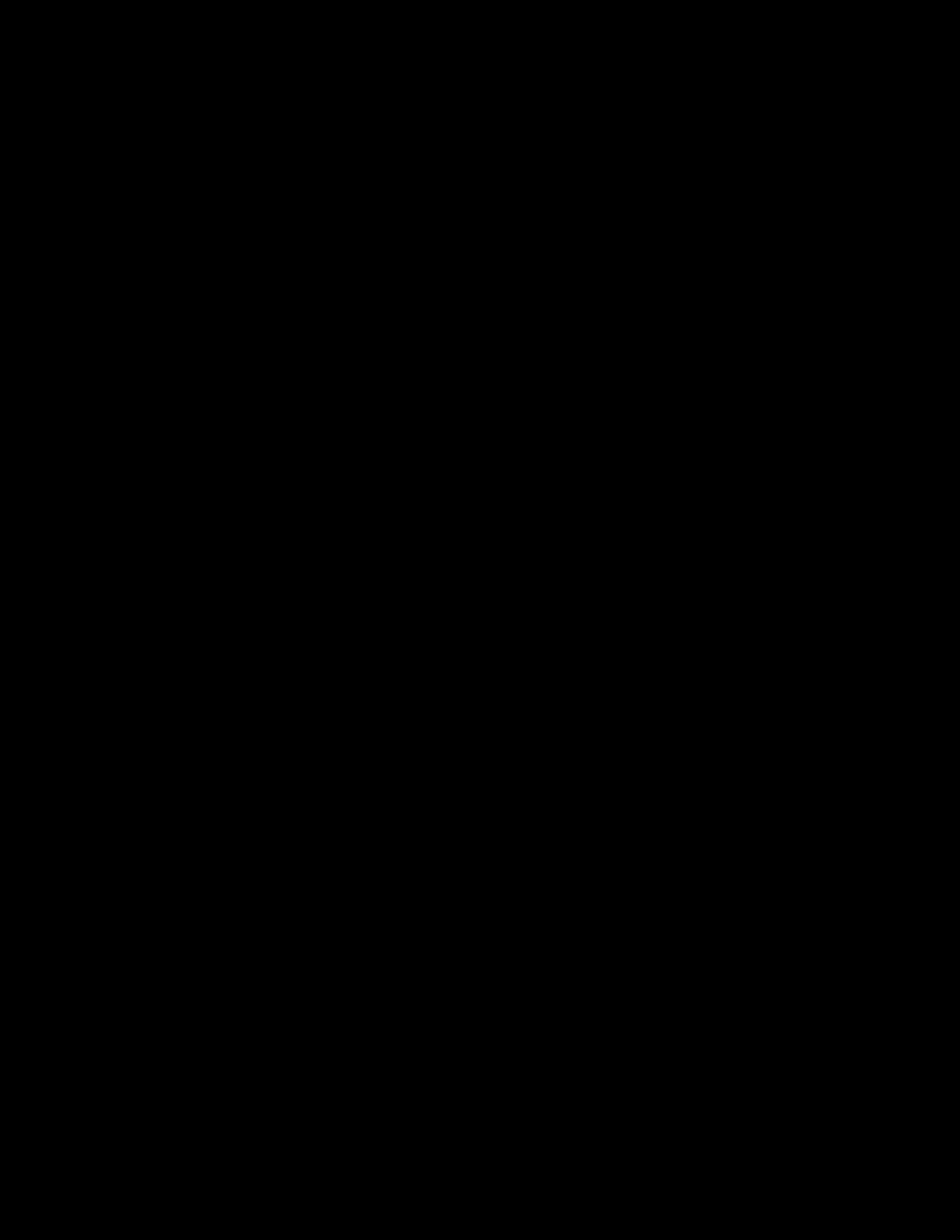}
      & 
       \imagepaste{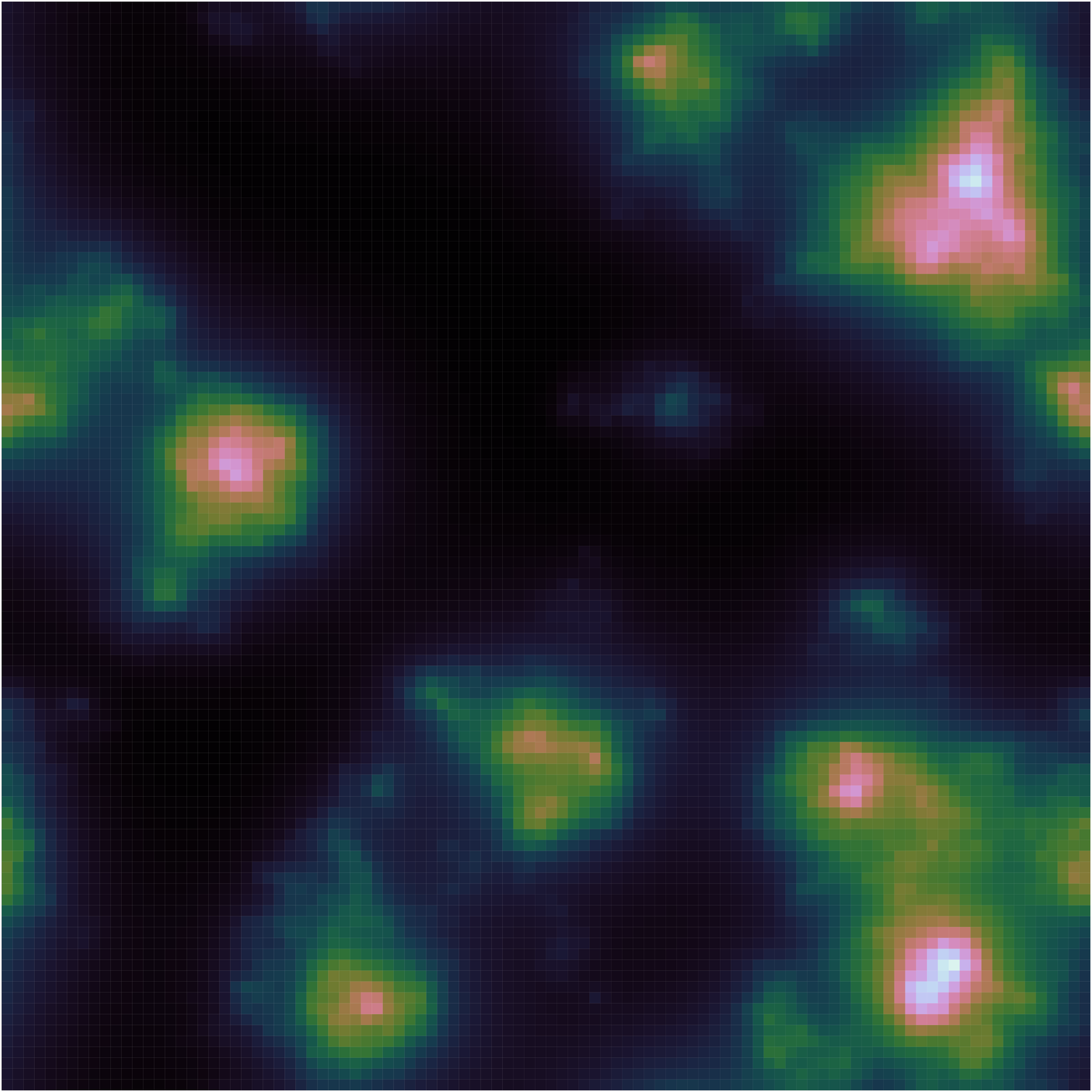}
      \\ \bottomrule
      \vspace{-2.5cm}  
       \begin{tabular}{p{0.15cm} p{1.5cm}}
       (b) & $\hat{\xi}_{\rm LS}(r)$
       \end{tabular}
       &
      \graphpaste{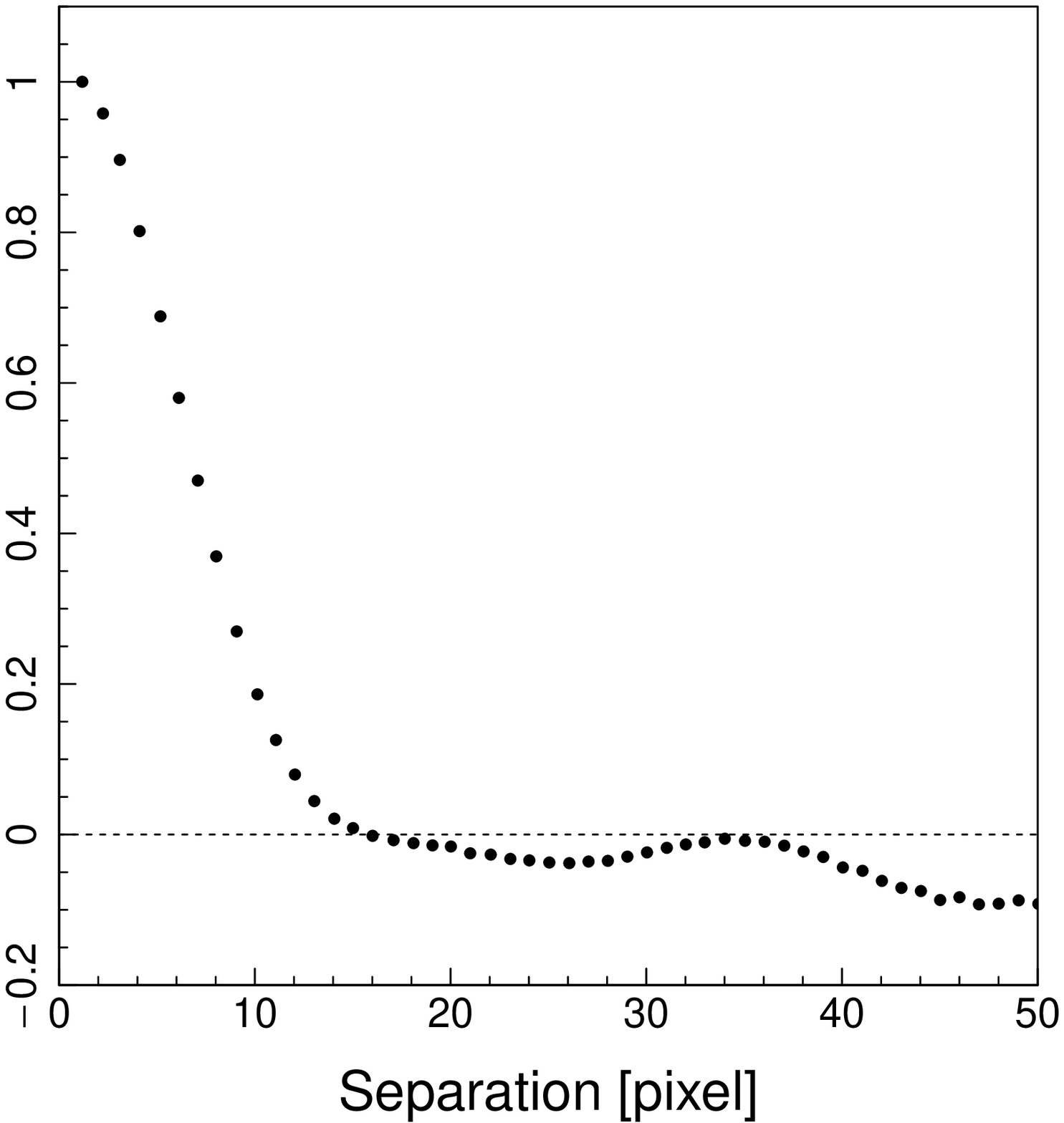}
      & 
      \graphpaste{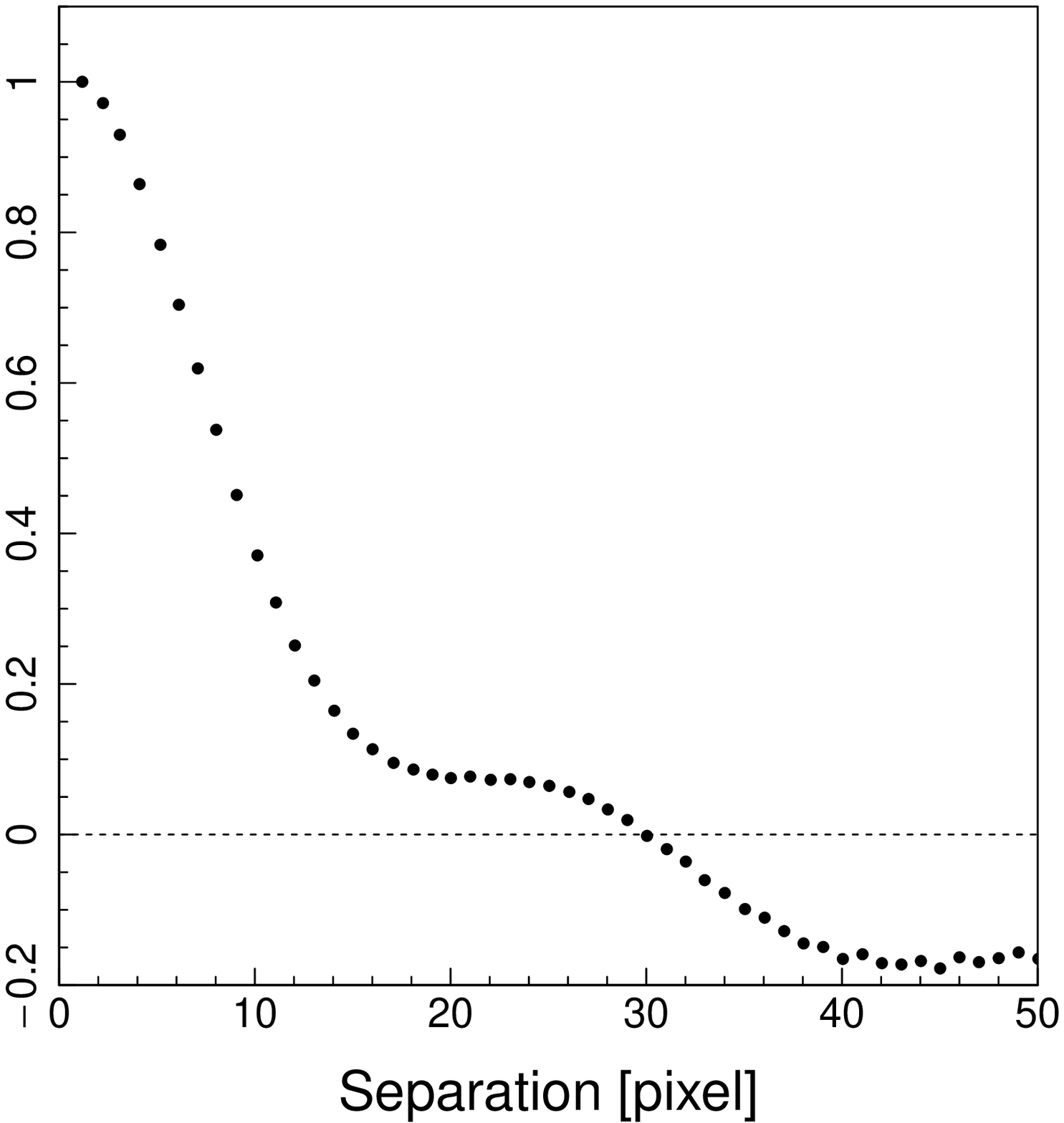}
      & 
      \graphpaste{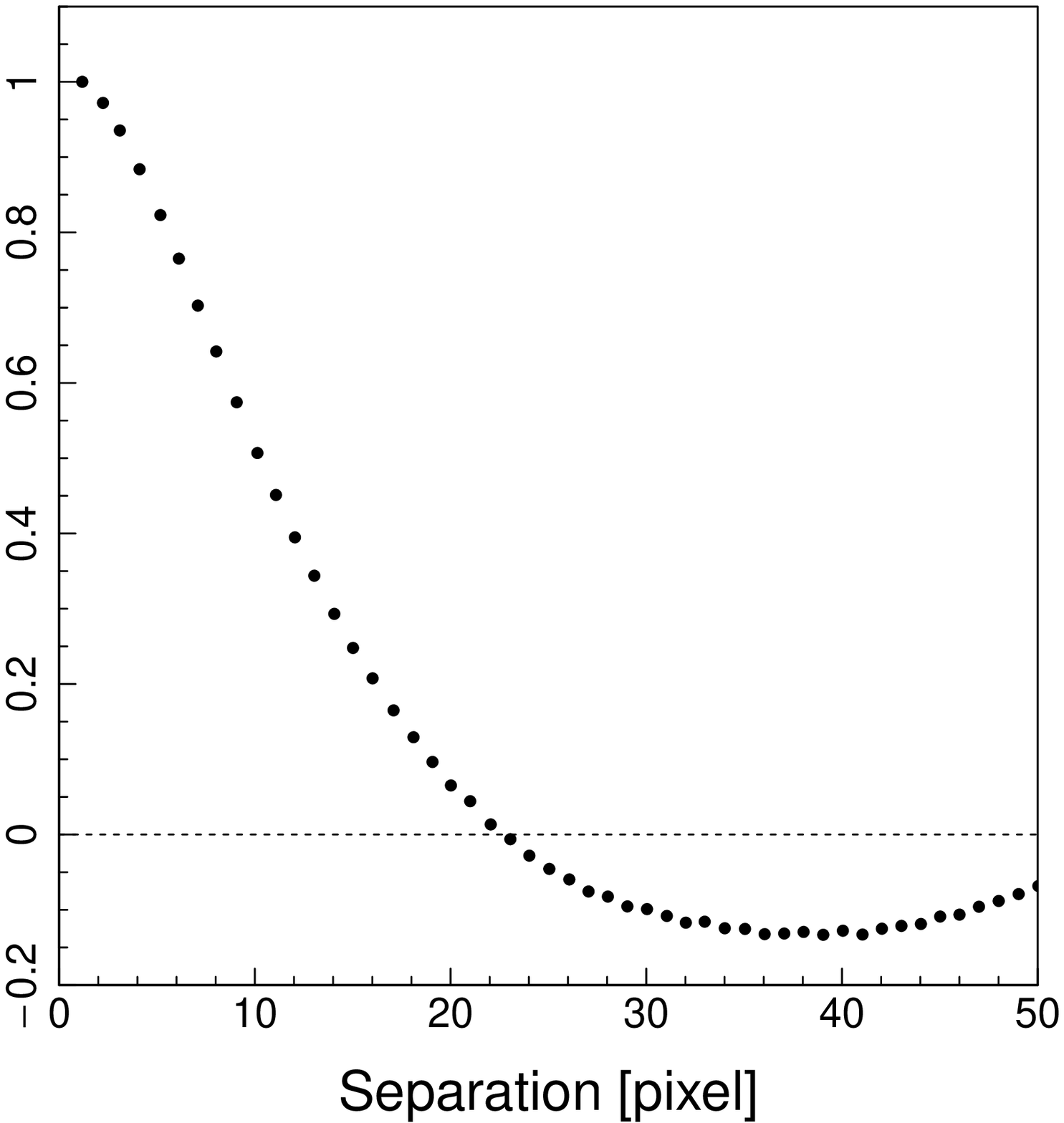}
      \\ \bottomrule
      
            \vspace{-2.5cm}  
       \begin{tabular}{p{0.15cm} p{1.5cm}}
       (c) & $\sqrt{r}\tilde{\xi}(r)$
       \end{tabular}
       &
      \graphpaste{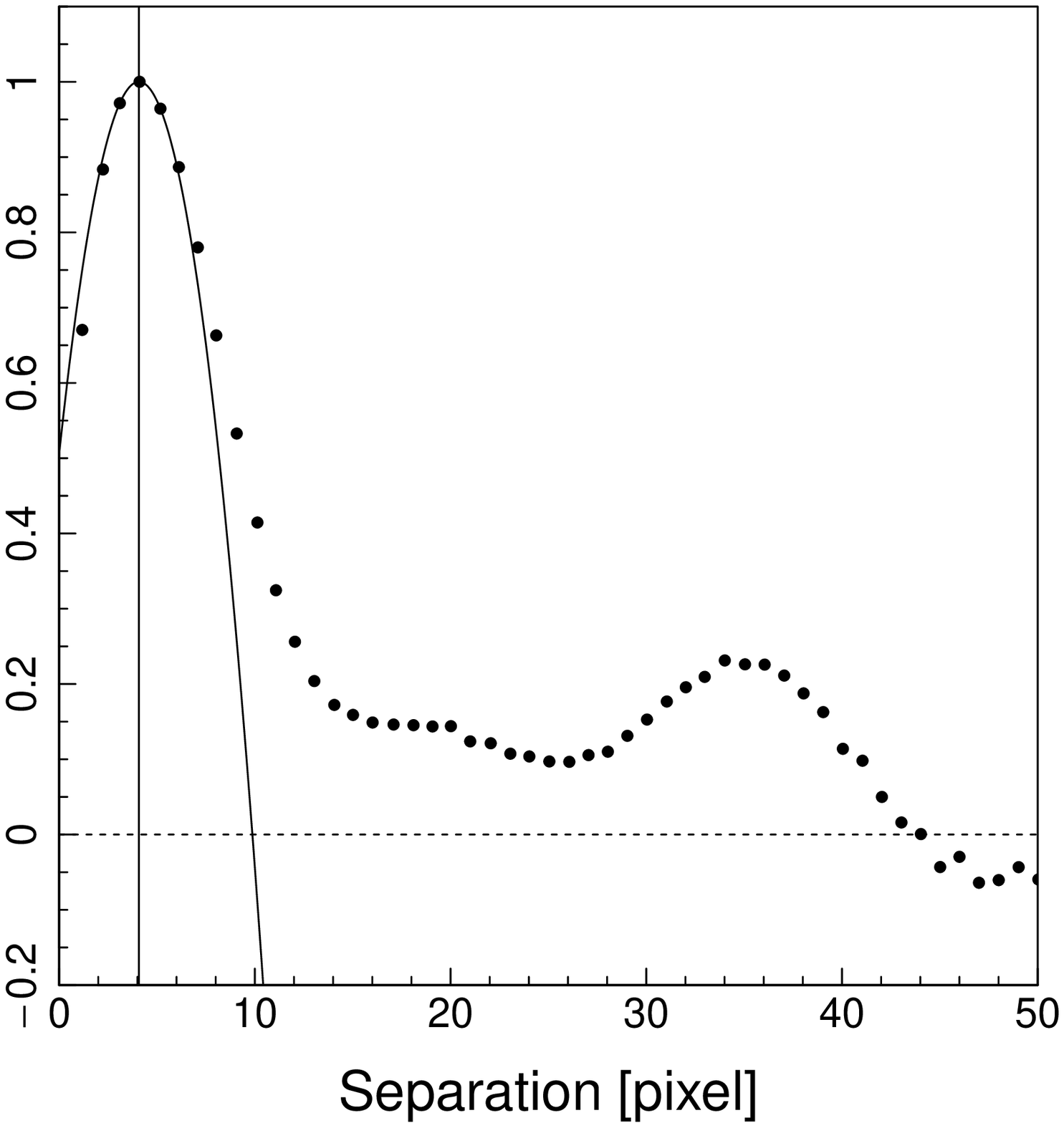}
      & 
      \graphpaste{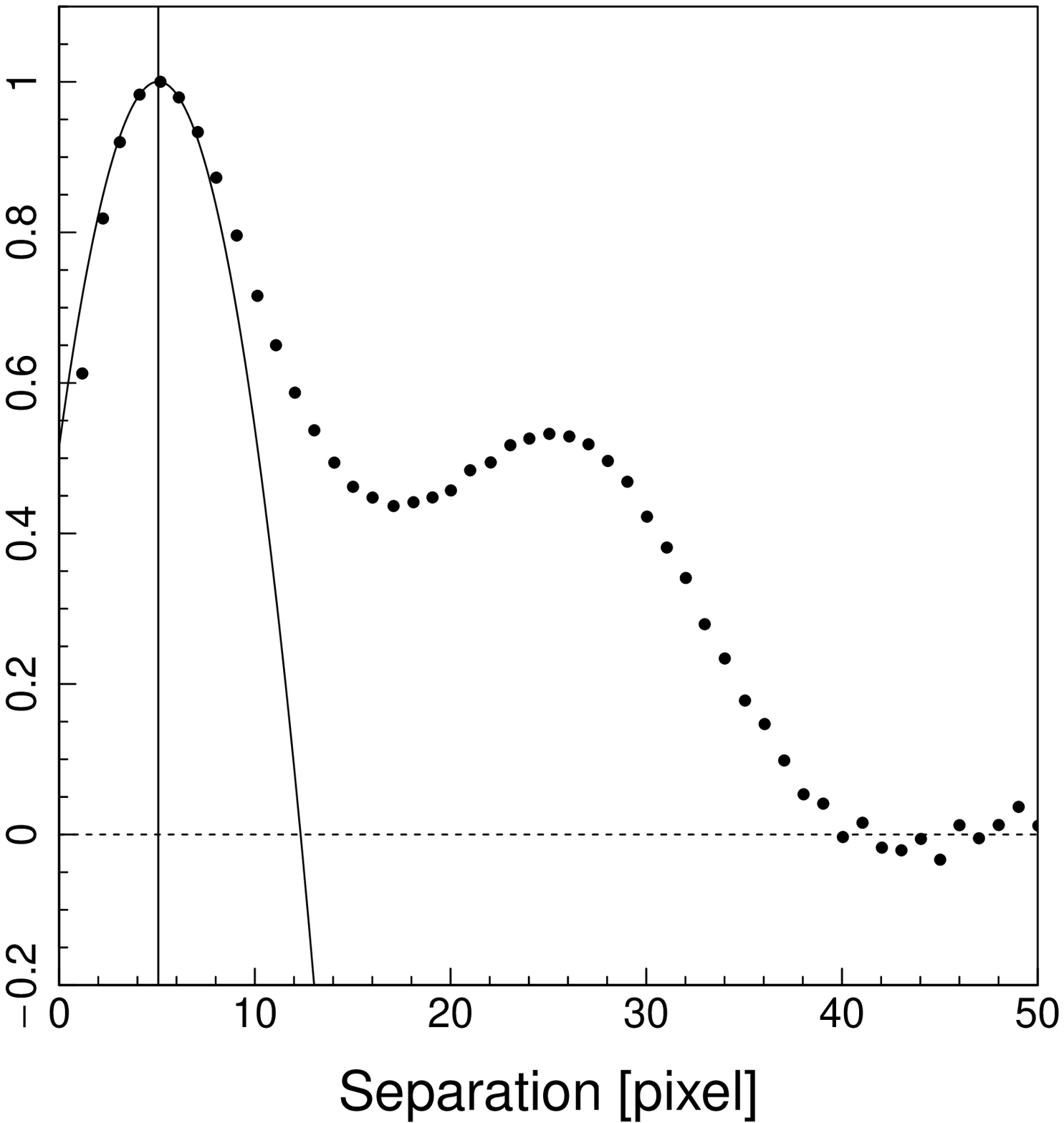}
      & 
      \graphpaste{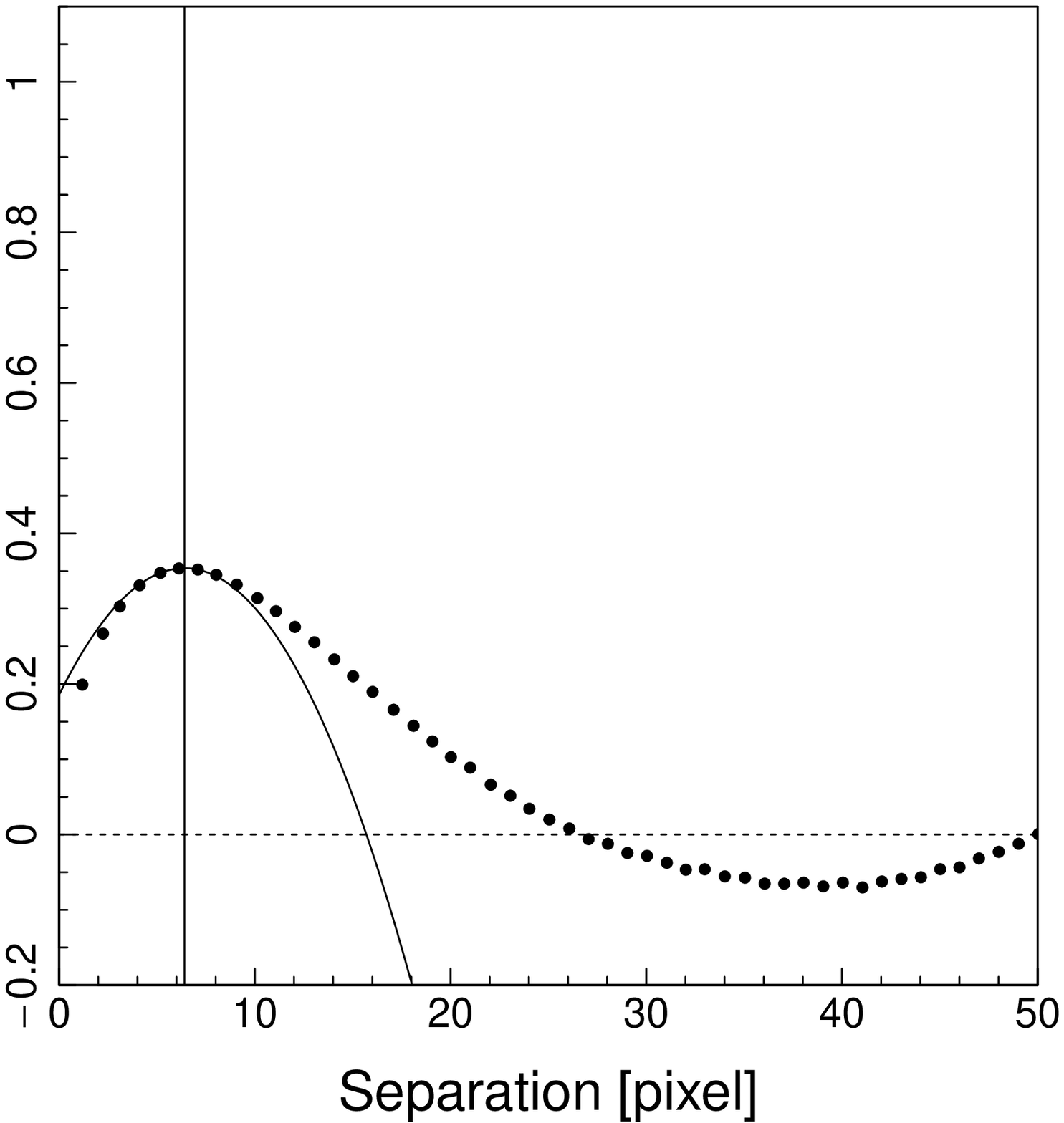}
      \\ \bottomrule
      
      \vspace{-2.5cm}
       \begin{tabular}{p{0.15cm} p{1.5cm}}
       (d) & Recovered clump size
       \end{tabular}
       &
      \graphpaste{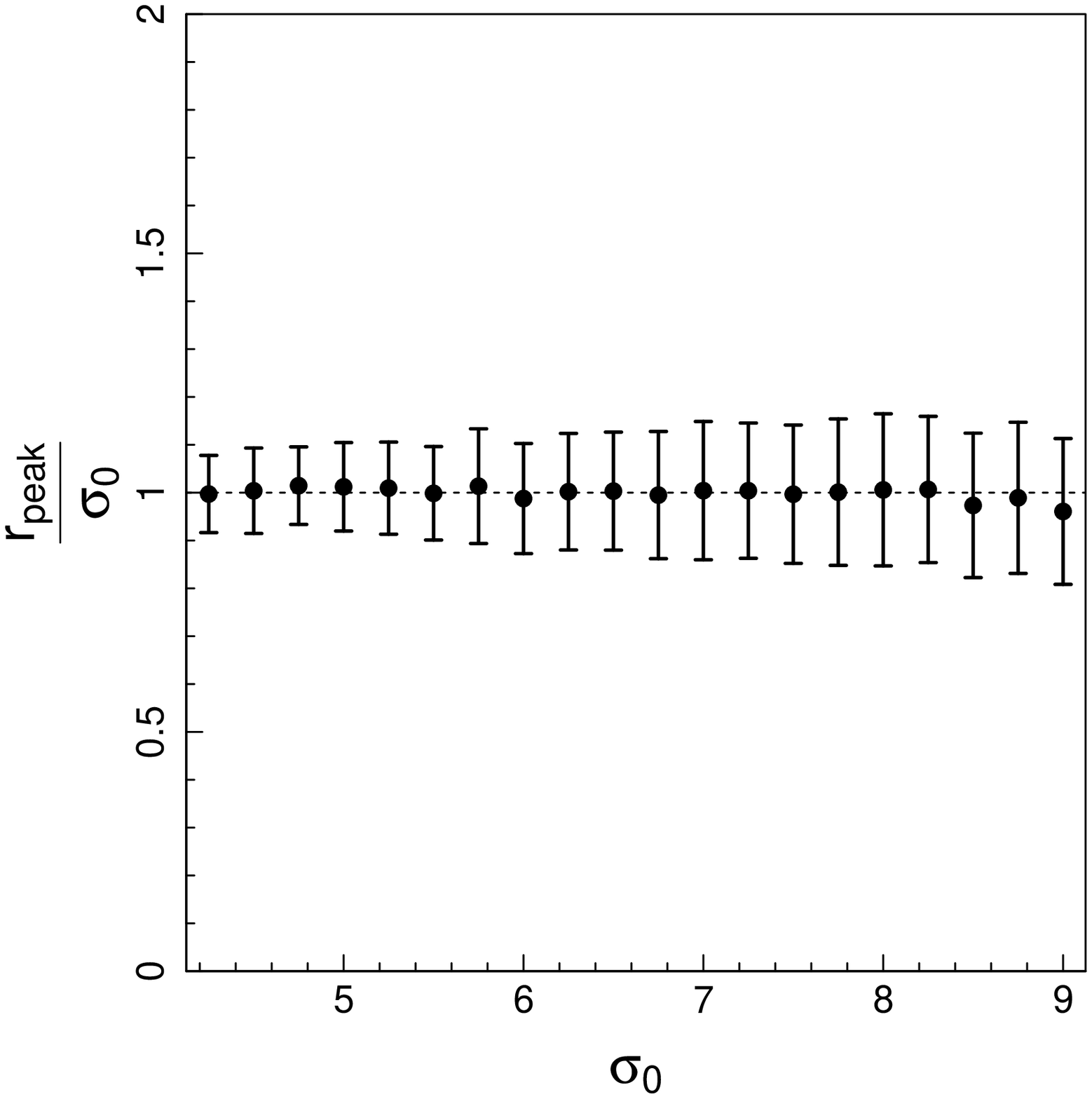}
      & 
      \graphpaste{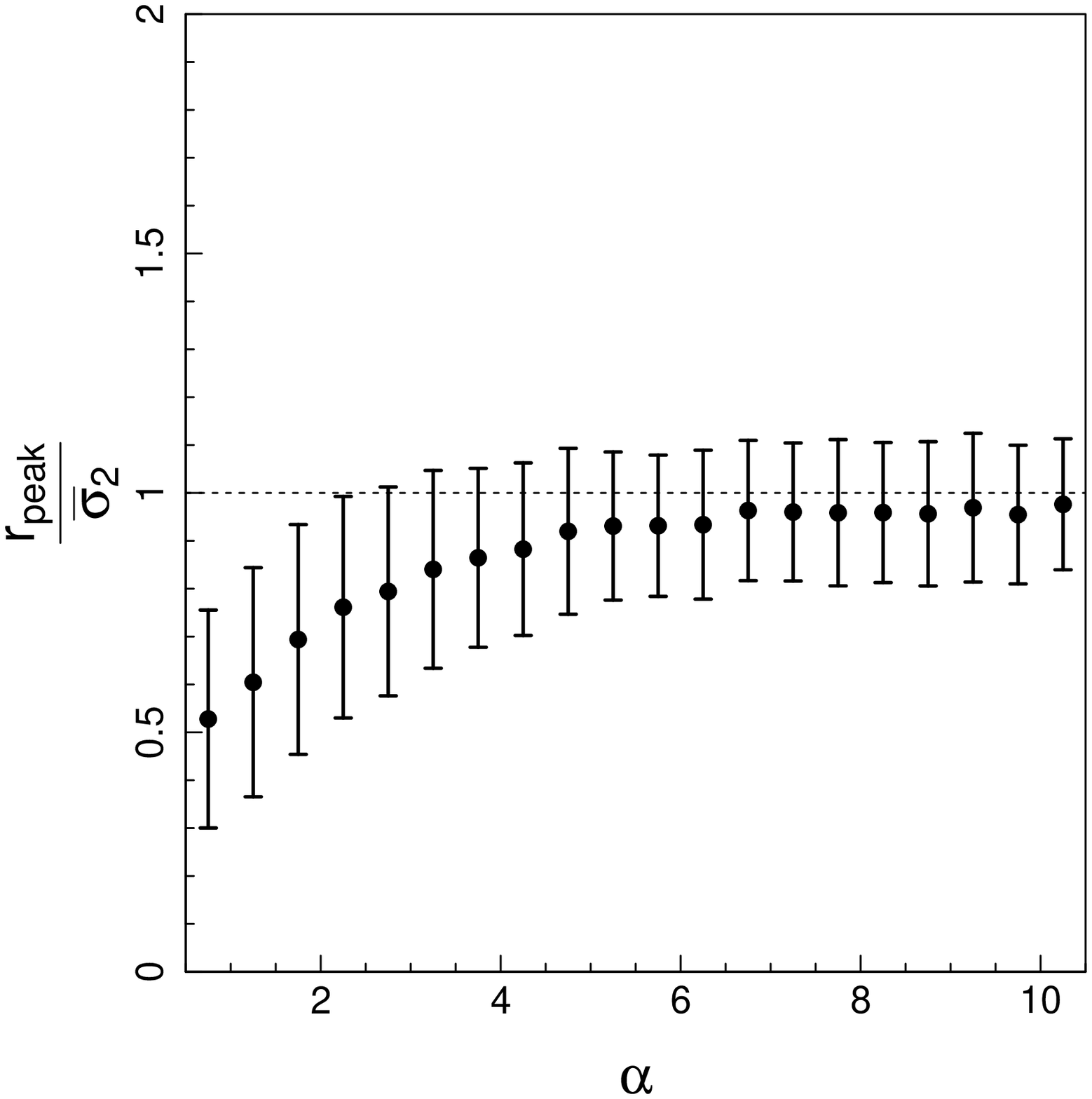}
      & 
      { \includegraphics[trim=0 0 0.5 -0.5cm, clip=true,width=0.23\textwidth]{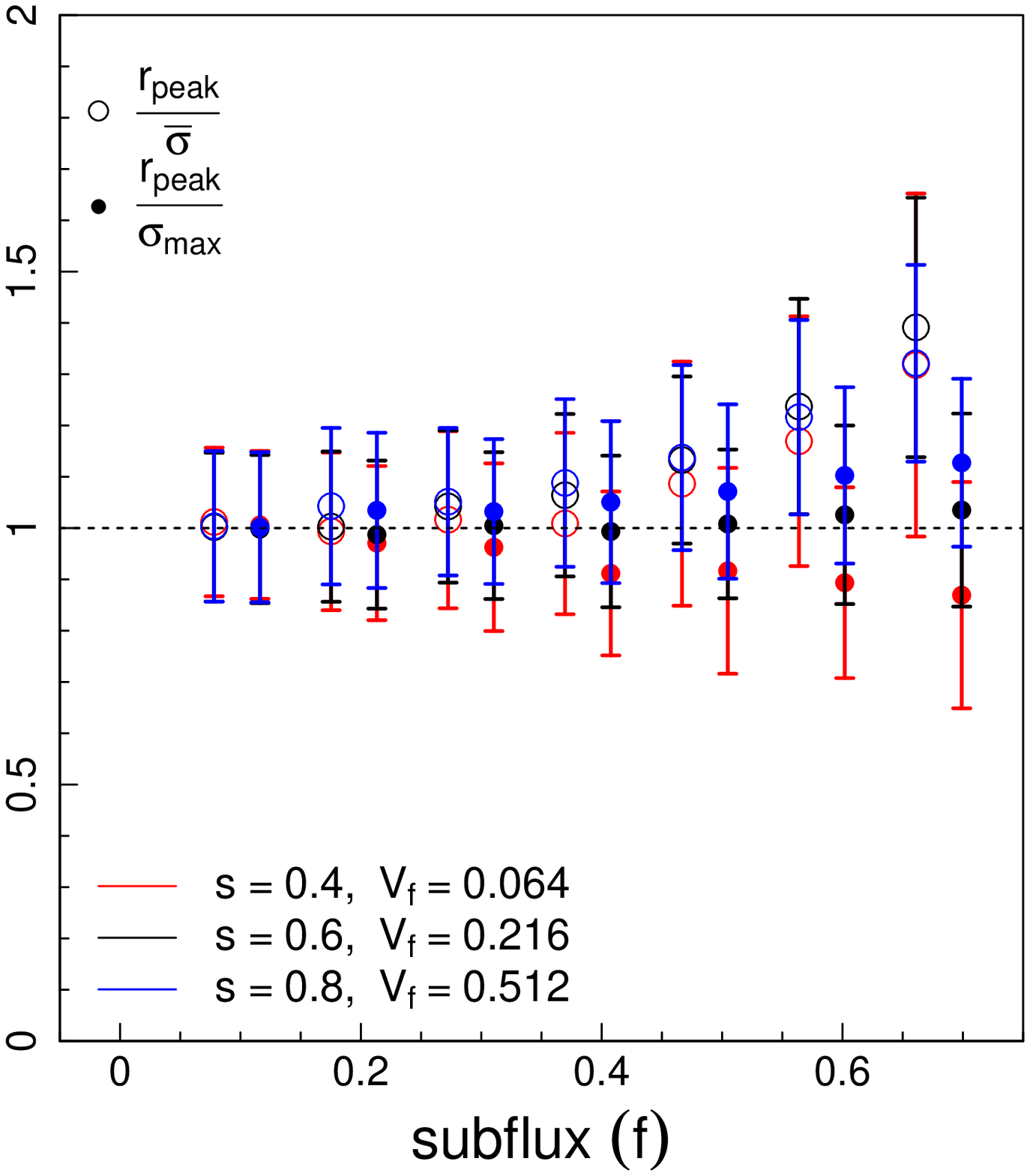} }
      \\ \bottomrule
      
      \end{tabular}
      \caption{Summary of the three clump models explored in Section \ref{sec:statistics}. The first three rows show the 2D mock data, the 2PCF and the $\mathit{w}$2PF of a single realization with parameters $\sigma_{0}=4$, $\alpha=3$ and $f=s=0.5$. The solid curve in row (c) is a parabolic fit to the maximum point and its neighbors on either side, used to determine the maximum position $r_{\rm peak}$ (vertical solid line) at sub-pixel level. The final row shows average of $r_{\rm peak}$ and its standard deviation for 250 random realizations for each clump model, as a function of the characteristic parameters of this model (details in Section \ref{subsec:SS}). In the clump model with substructure (third column), the parameters $f$ and $s$ denote the fractional flux and size of substructure relative to its parent structure (recursively). $V_f\equiv s^3$ denotes the volume fraction of substructure.}
      \label{tbl:distribution}
      \end{center}
      \end{figure*}
      
  \subsection{Effect of Gaussian blurring} \label{subsec:blur}
  
We wish to quantify the effect of blurring on the recovery of the primary clump size using the $\mathit{w}$2PF. A single Gaussian clump with size $\sigma_0$ convolved with a Gaussian of standard deviation $\sigma_{\rm blur}$ results in a Gaussian 2PCF with standard deviation 
  \begin{equation}\label{eq:blur}
  \sigma_{\rm final}=\sqrt{\sigma_0^2+\sigma_{\rm blur}^2}.
  \end{equation}
Consequently, the maximum position $r_{\rm peak, blur}$ of the $\mathit{w}$2PF is equal to $\sigma_{\rm final}$, and we can invert the relation to recover the maximum without blurring,
\begin{equation}
  r_{\rm \rm recovered}=\sqrt{r_{\rm peak, blur}^2-\sigma_{\rm blur}^2}.
\end{equation}
  
  If a density field is made of multiple randomly placed and equally sized Gaussian clumps, Eq.~(\ref{eq:blur}) is not strictly true, because of clump-clump correlations. However, since these correlations are random, Eq.~(\ref{eq:blur}) still remains true for an ensemble of fields. Or, equivalently, $r_{\rm recovered}$ is the expectation of the clump radius $r_{\rm peak}$. For more complicated density fields, such as the power-law and substructure models in Table \ref{tbl:distribution}, the Gaussianity of the 2PCF no longer holds, not even in an ensemble sense. Hence, we expect the $r_{\rm \rm recovered}$ to systematically differ from the true unblurred measurement $r_{\rm peak}$.
  
\begin{figure*} 
\centering
\begin{tabular}{ccc}
	\hspace*{-0.5cm} \includegraphics[trim=0 0 0 0 clip=true,width=0.35\textwidth]{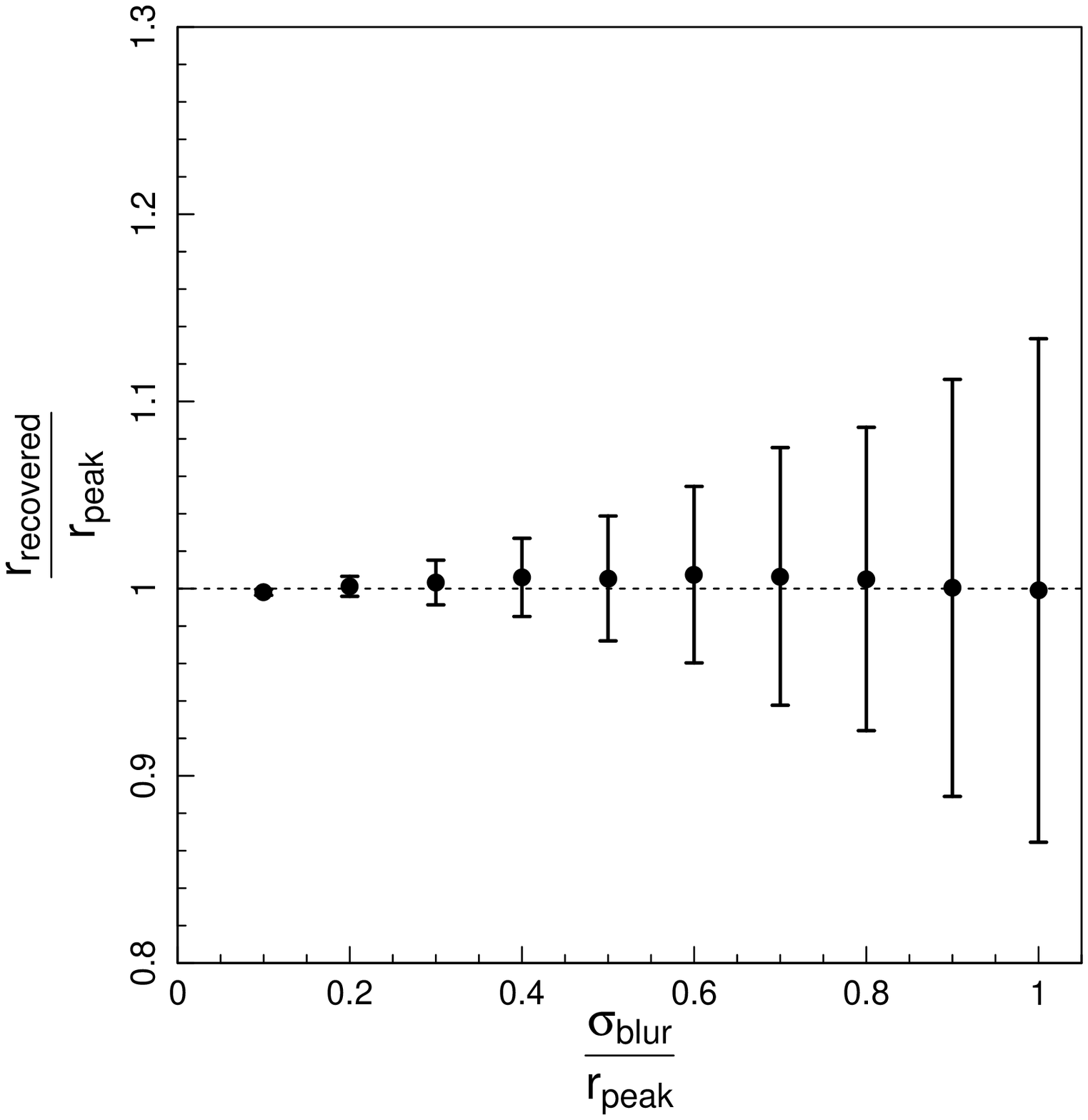}  &
	\hspace*{-0.25cm} \includegraphics[width=0.35\textwidth]{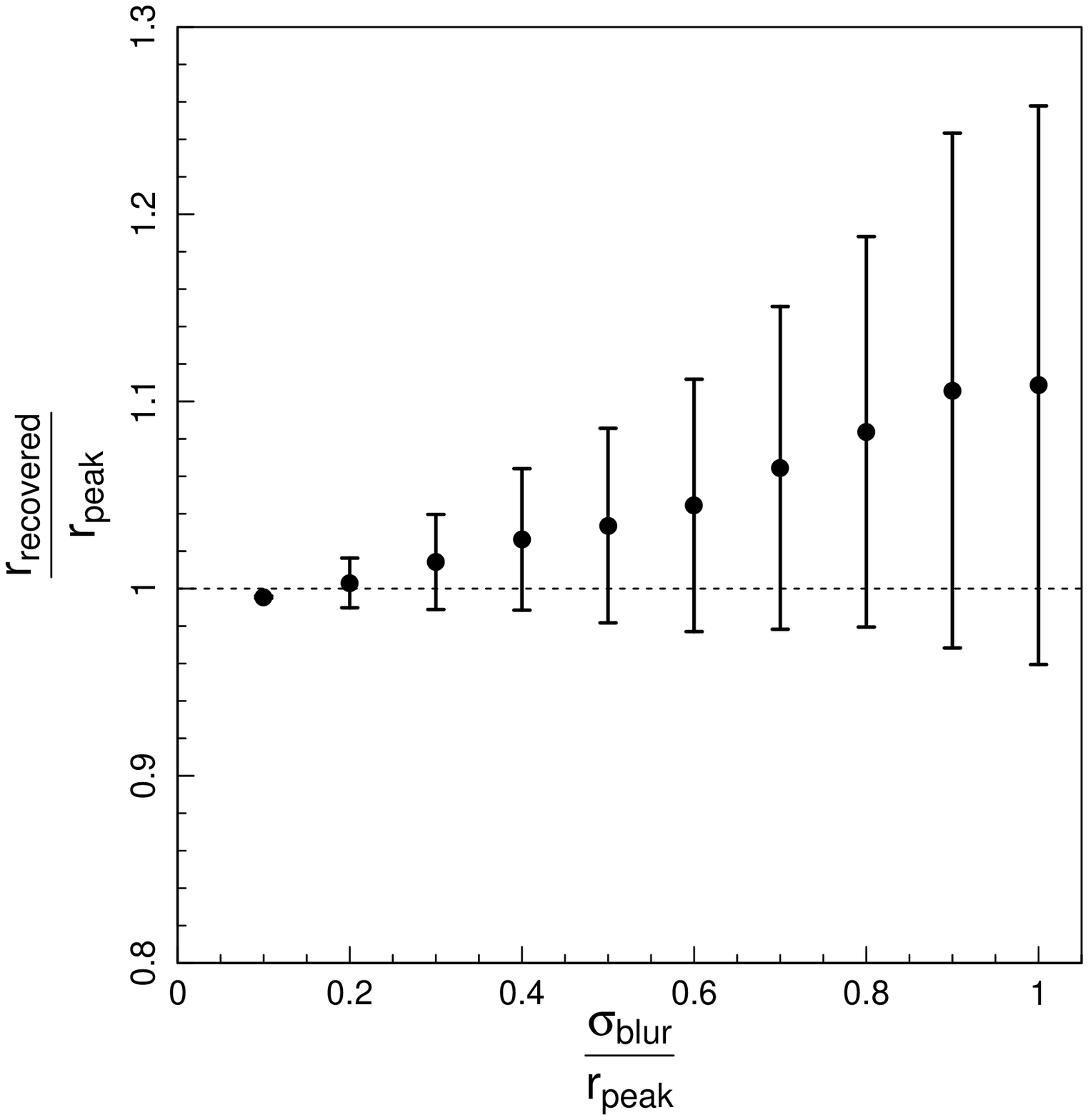} &
	\hspace*{-0.25cm} \includegraphics[width=0.35\textwidth]{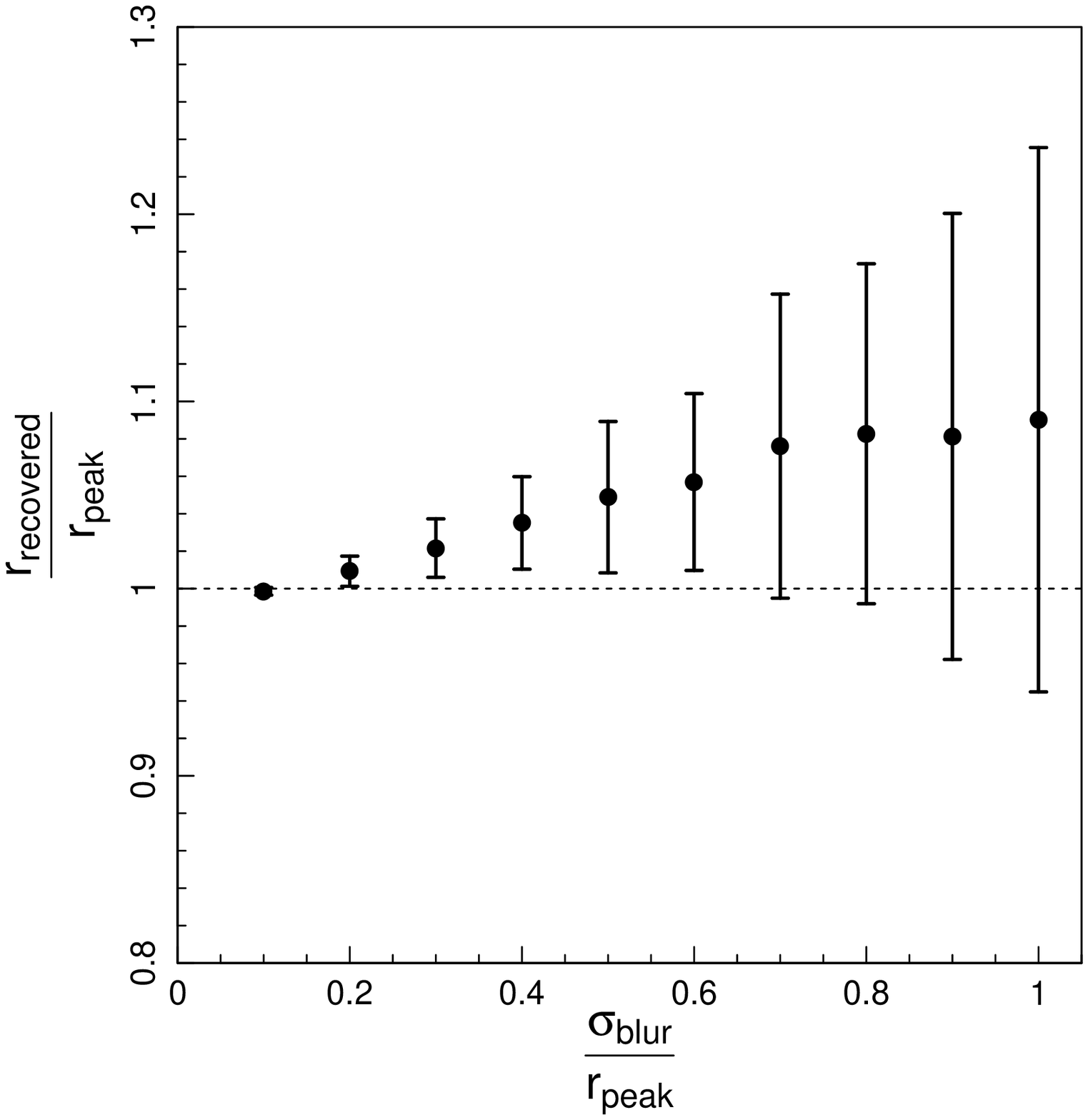} 
\end{tabular}
\caption{Sensitivity of the recovered clump size, $r_{\rm recovered}$, to the standard deviation of the convolution parameter, $\sigma_{\rm blur}$, of a Gaussian PSF, relative to the input clump size $r_{\rm peak}$. The three panels show the three clump models of Table \ref{tbl:distribution}. The points and error bars indicate mean and 1$\sigma$ of 400 random realizations.}
\label{fig:modelblur}
\end{figure*}

  To quantify the relation between $r_{\rm \rm recovered}$ and $r_{\rm peak}$, we reuse the mock data fields shown in Table \ref{tbl:distribution} (for parameters $\sigma_0=4$, $\alpha=3$, $s=0.5$, $f=0.5$), but blurred by a Gaussian kernel of standard deviation $\sigma_{\rm blur}$. The mean and standard deviation of 400 random realizations for each blurring size are shown in Figure \ref{fig:modelblur}.
    
We find for sufficiently small blurring sizes ($\leq 20\%$ of clump sizes) that the uncertainty due to blurring is negligible, $<1\%$. However, there is a systematic effect on $r_{\rm peak}$ measurements of power-law and substructure models which asymptote to $\approx10\%$ for blurring sizes similar to size of the primary clump -- still an acceptable error in most practical cases. 

In the case of DYNAMO-HST data analyzed in Section \ref{sec:DYNAMO}, the PSF size is $\leq30\%$ which, using a conservative estimate, adds a systematic effect of $+1\%$ and an uncertainty of $\pm 1\%$ to the final value. We take this into account when estimating $r_{\rm peak}$ of the DYNAMO-HST maps.

   \subsection{Effect of Gaussian White noise} \label{subsec:noise}
   
Another important factor affecting the clump size measurements is the image noise. We only consider the noise in the $D$-field, since this largely dominates over the noise in the $R$-field, because the $D$-field is typically based on emission line maps, whereas $R$-fields are based on continuum maps, spanning a much larger range in wavelength. Conventional methods identify clumps as structures above a fixed threshold over the RMS noise. Hence reducing the noise level leads to measuring either larger sizes or more clumps of smaller size. In contrast, we expect our statistical method to show much less systematic variation with noise. To test this claim and measure the statistical uncertainty caused by image noise, we contaminate our mock images by random noise. As in the previous section we run 400 random realizations. To each $D$-field we add Gaussian pixel noise of standard deviation $\sigma_{\rm noise}$ and then compare the $r_{\rm peak}$ values extracted from the $\mathit{w}$2PF.

To quantify the noise level in a resolution-independent manner, we choose the following definition: the noise $N$ is defined as the nearest-neighbor, standard deviation, which is simply the standard deviation of the difference in flux between pixel and its adjacent neighbor (in both dimensions). The `signal' $S$, on the other hand, is defined as the mean of the $5\%$ brightest pixels of the $D$-field. In this way, the relative noise $N/S$ is independent of the pixels' size in the case of Gaussian white noise. A note of caution: since pixel-to-pixel flux is correlated in HST images we should expect weak systematic variation under this definition.
   
\begin{figure*} 
\centering
\begin{tabular}{ccc}

	\hspace*{-0.35cm} \begin{tikzpicture}[      
        			every node/.style={anchor=south west,inner sep=0pt},
        			x=1mm, y=1mm,
      			]   
      	\node (fig1) at (0,0)
			{ \hspace*{-0.25cm} \includegraphics[width=0.35\textwidth]{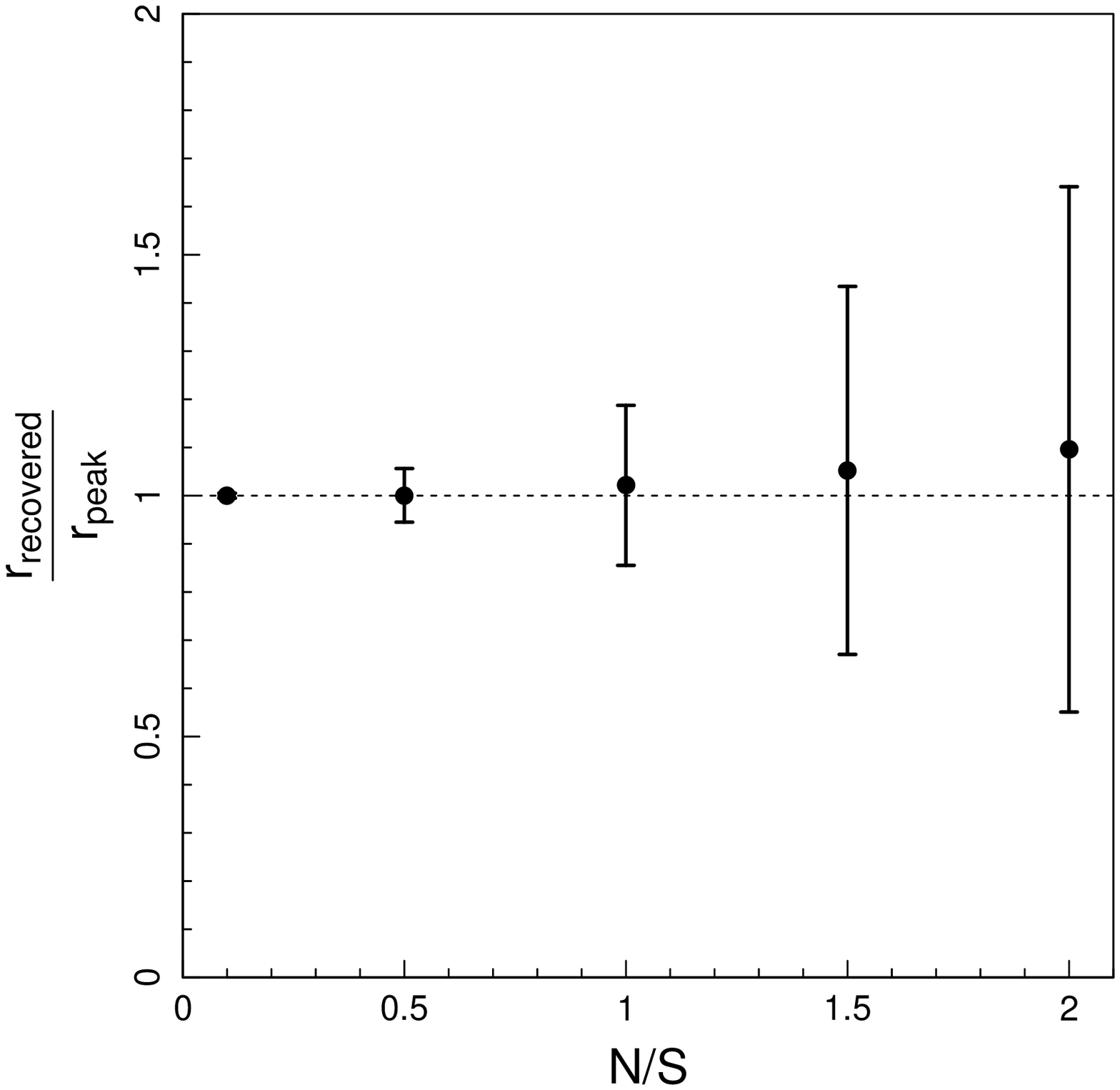} };
	\node (fig2) at (8.77,6)
			{ \includegraphics[trim=0 0 0 0 clip=true,width=0.075\textwidth]{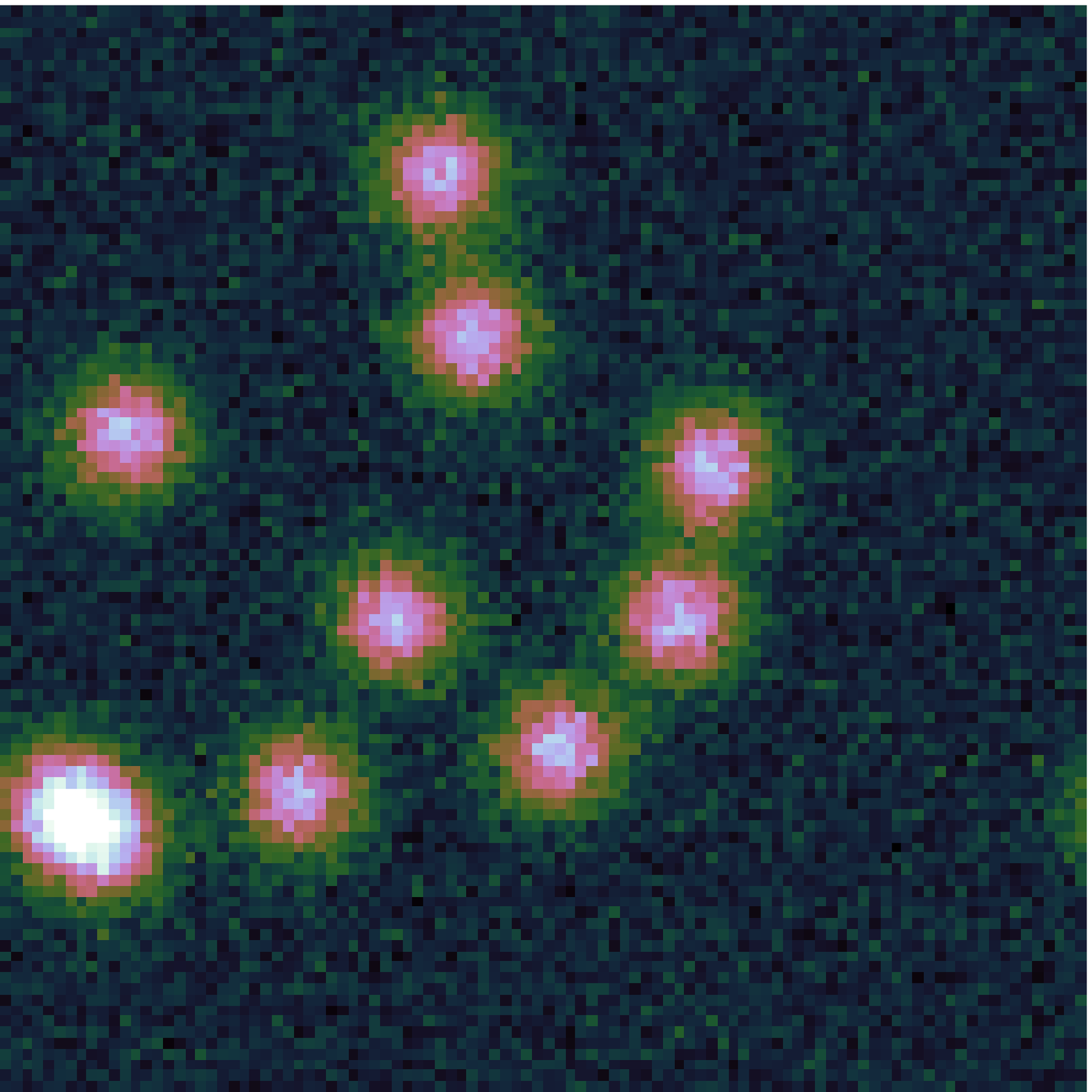} };
	\node (fig3) at (27.3,6)
			{ \includegraphics[trim=0 0 0 0 clip=true,width=0.075\textwidth]{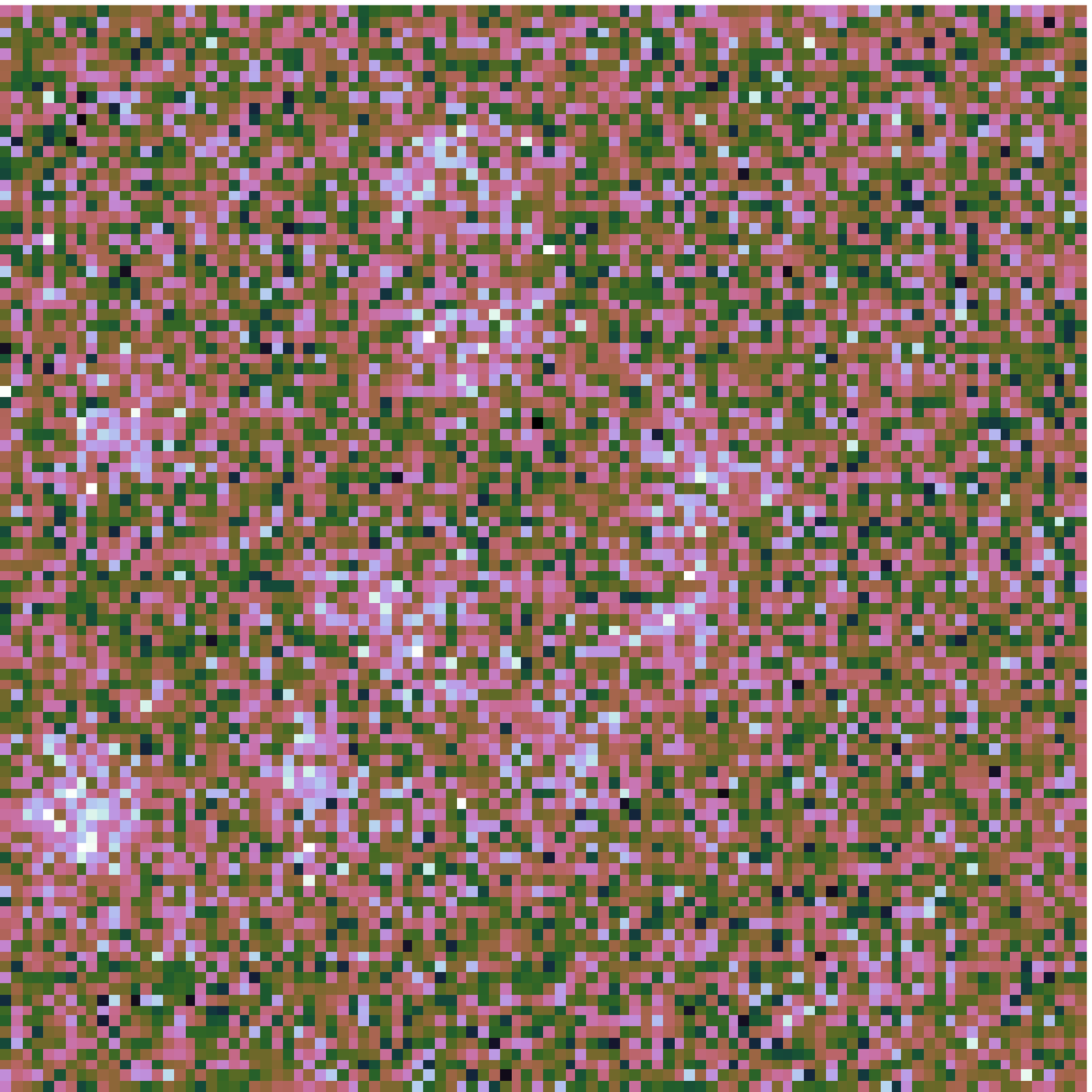} };
	\node (fig3) at (48.1,6)
			{ \includegraphics[trim=0 0 0 0 clip=true,width=0.075\textwidth]{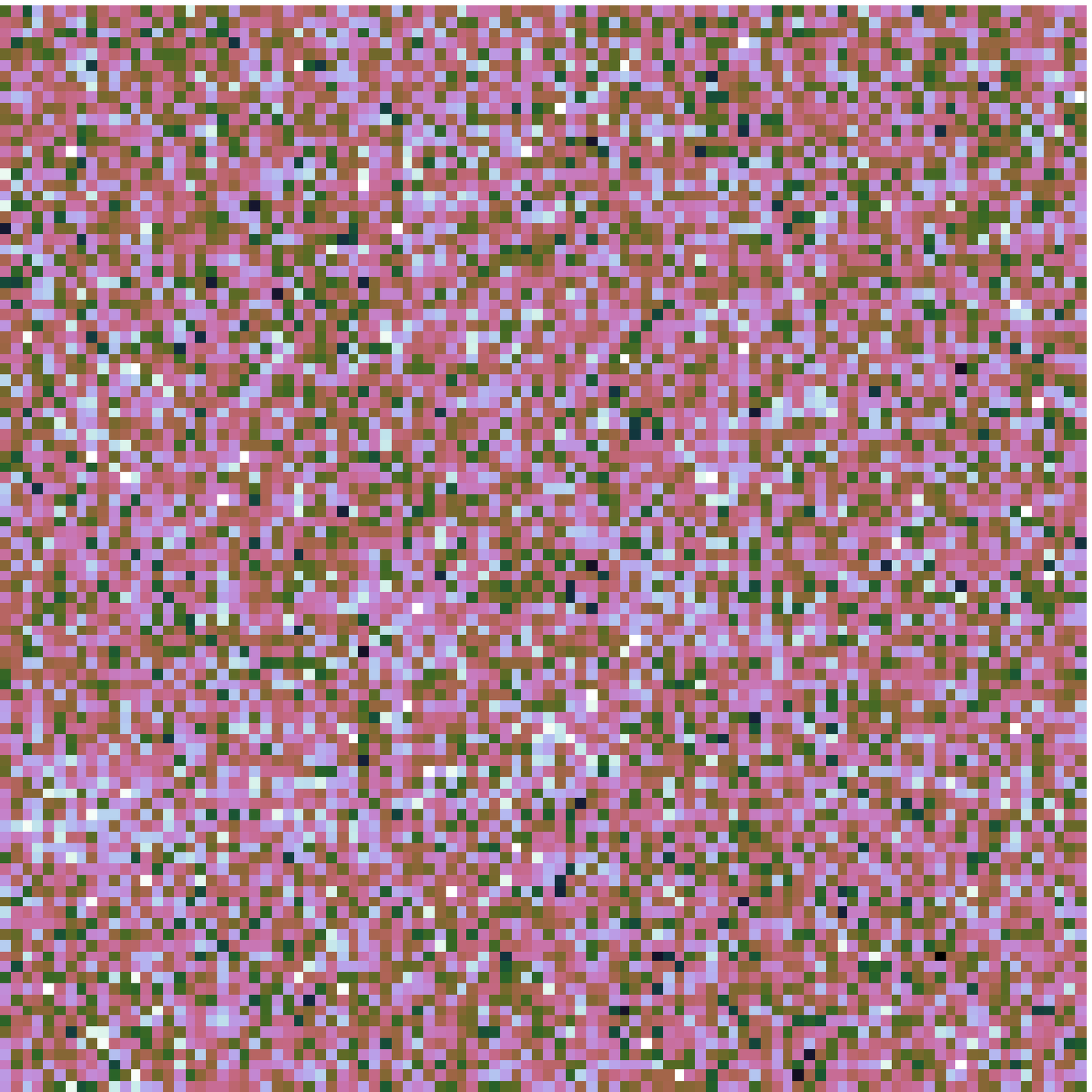} };
	\end{tikzpicture} &

	
	\hspace*{-0.25cm} \begin{tikzpicture}[      
        			every node/.style={anchor=south west,inner sep=0pt},
        			x=1mm, y=1mm,
      			]   
      	\node (fig1) at (0,0)
			{ \hspace*{-0.25cm} \includegraphics[width=0.35\textwidth]{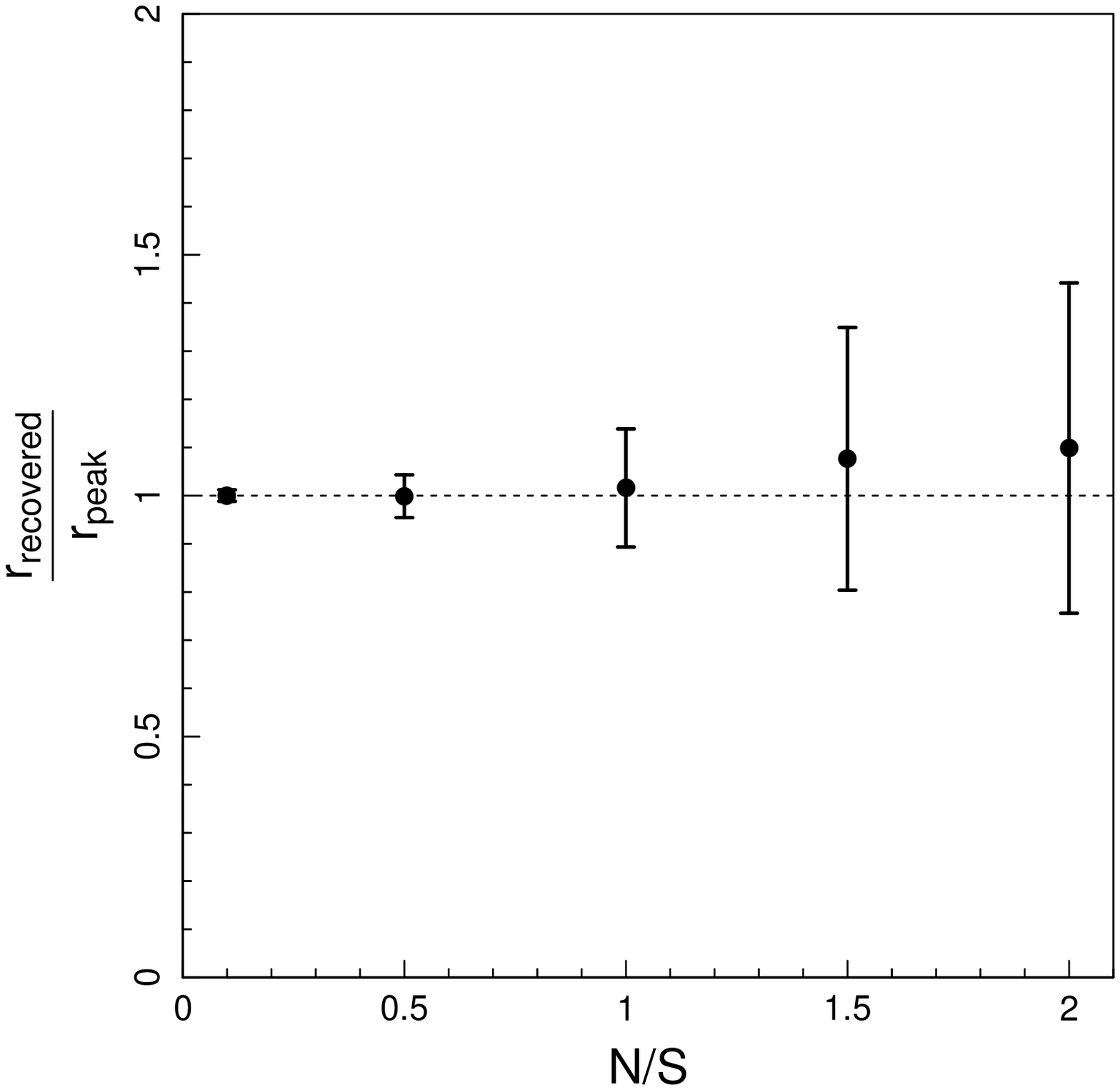} };
	\node (fig2) at (8.77,6)
			{ \includegraphics[trim=0 0 0 0 clip=true,width=0.075\textwidth]{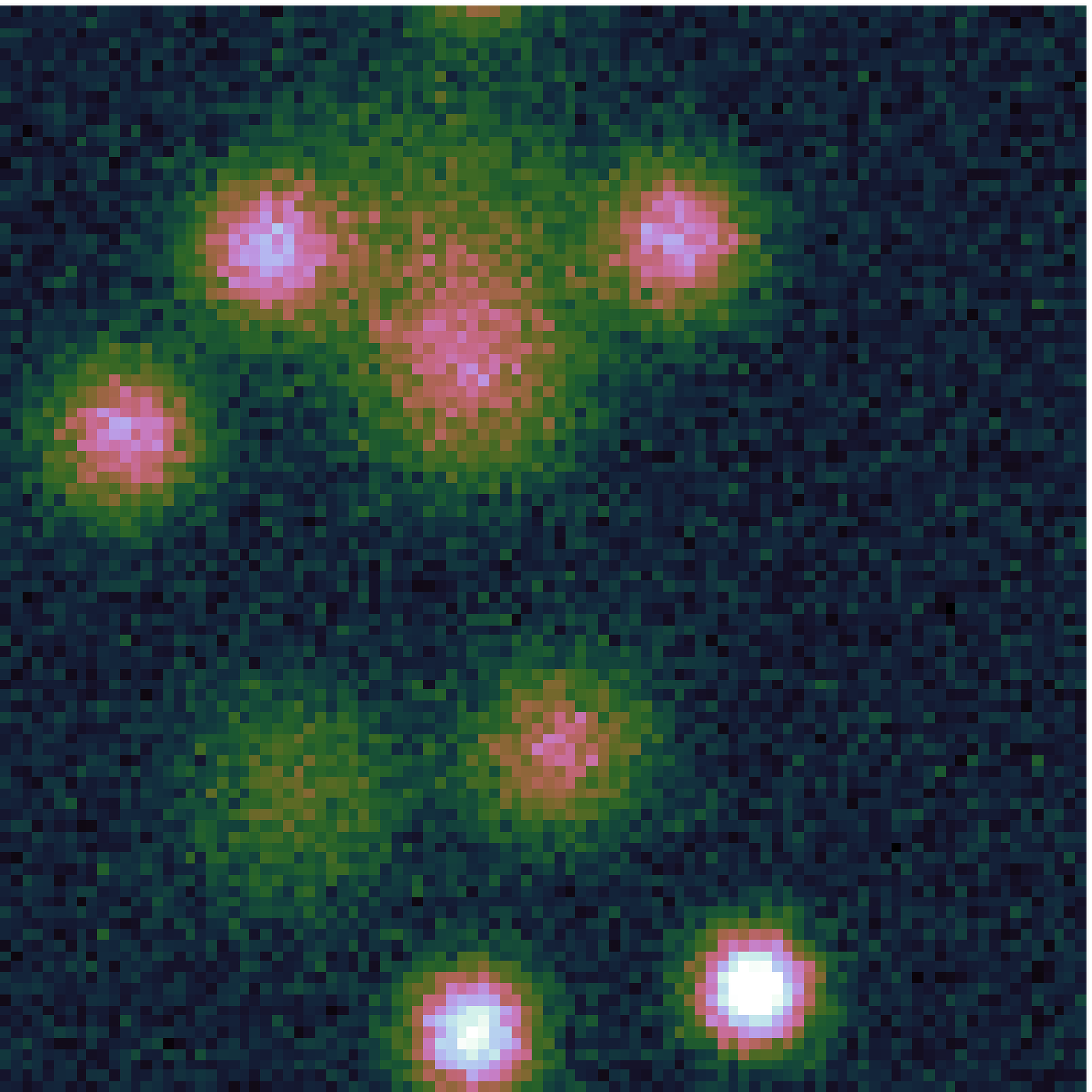} };
	\node (fig3) at (27.3,6)
			{ \includegraphics[trim=0 0 0 0 clip=true,width=0.075\textwidth]{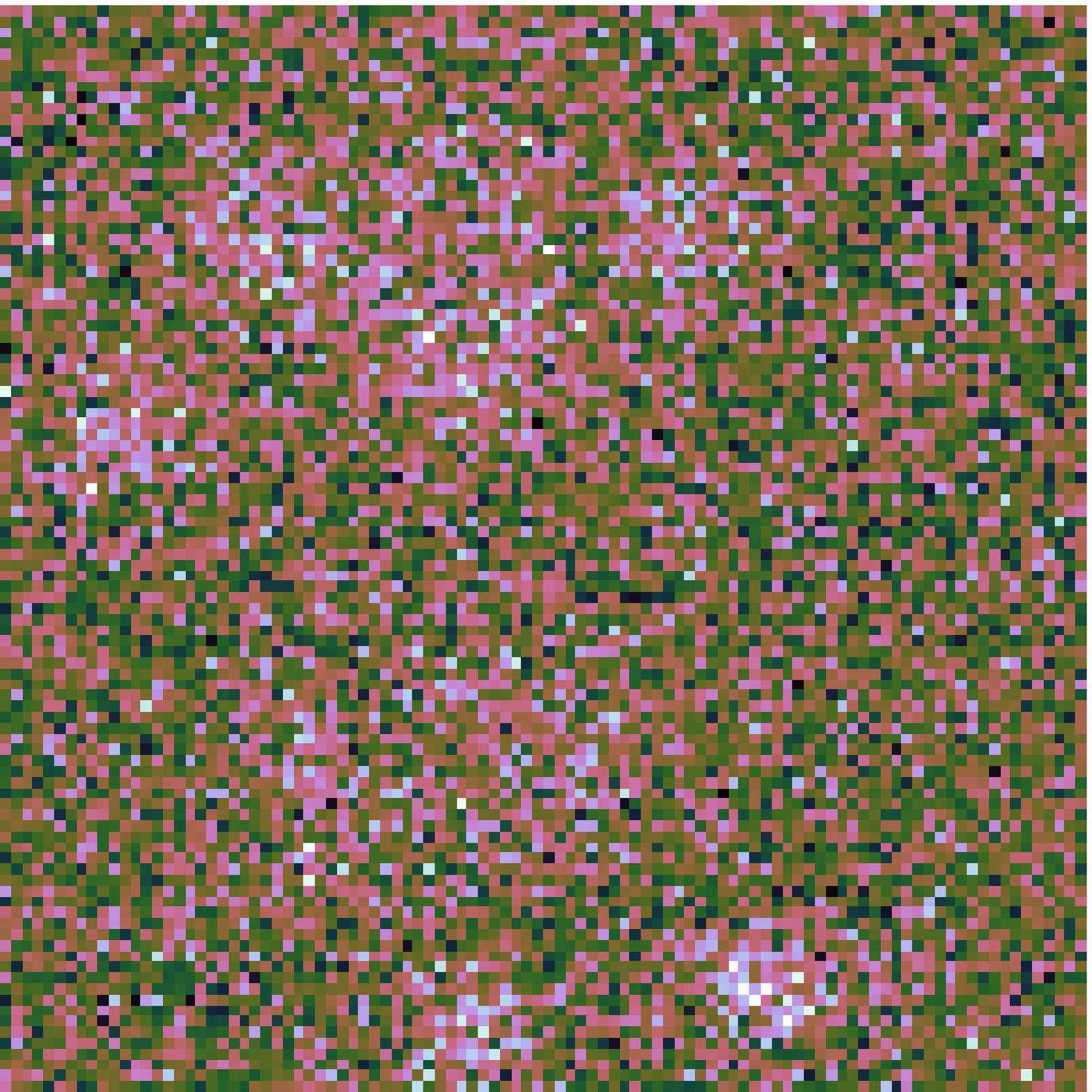} };
	\node (fig3) at (48.1,6)
			{ \includegraphics[trim=0 0 0 0 clip=true,width=0.075\textwidth]{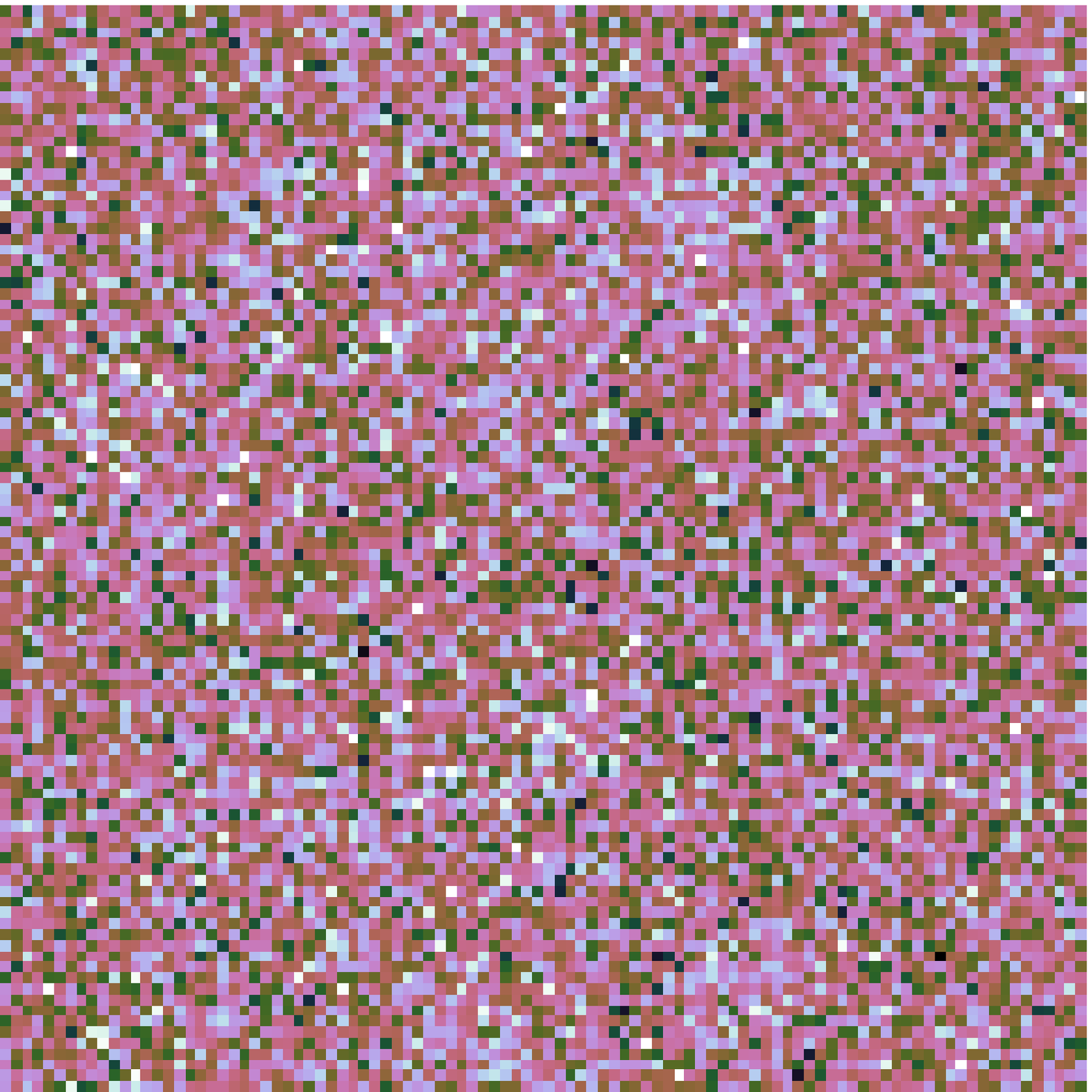} };
	\end{tikzpicture} &

	
	\hspace*{-0.25cm} \begin{tikzpicture}[      
        				    every node/.style={anchor=south west,inner sep=0pt},
        			x=1mm, y=1mm,
      				]   
      	\node (fig1) at (0,0)
			{ \hspace*{-0.25cm} \includegraphics[width=0.35\textwidth]{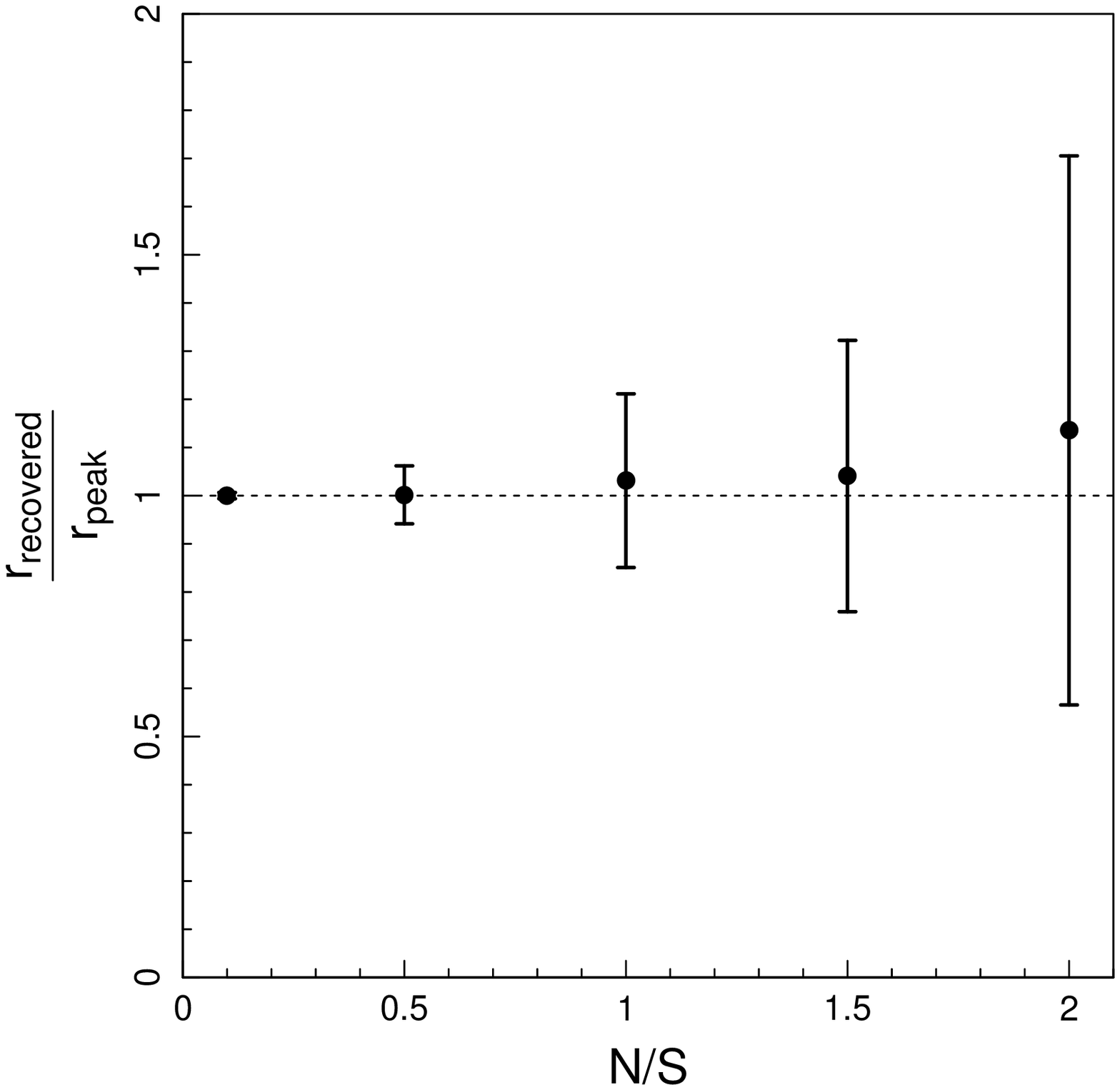} };
	\node (fig2) at (8.77,6)
			{ \includegraphics[trim=0 0 0 0 clip=true,width=0.075\textwidth]{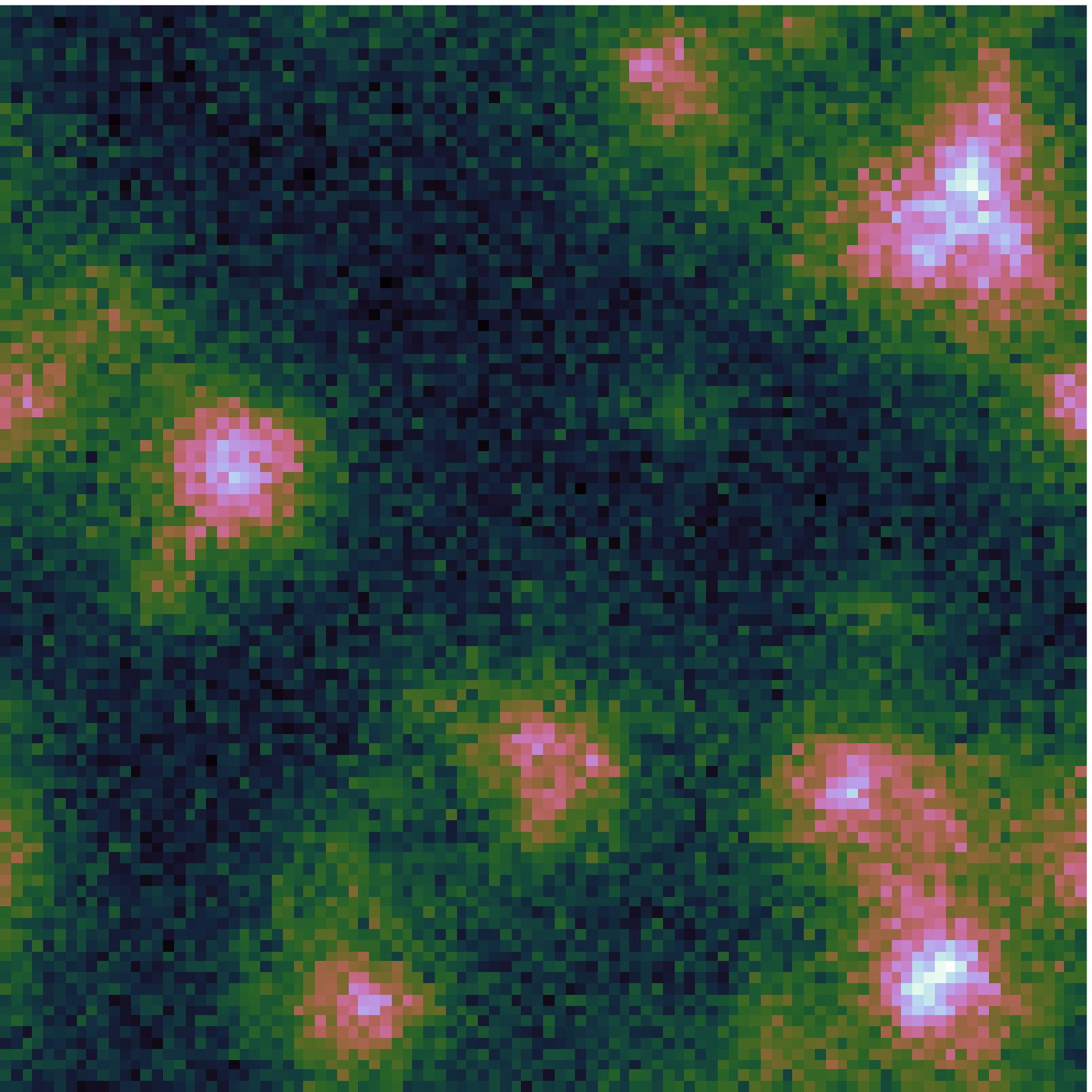} };
	\node (fig3) at (27.3,6)
			{ \includegraphics[trim=0 0 0 0 clip=true,width=0.075\textwidth]{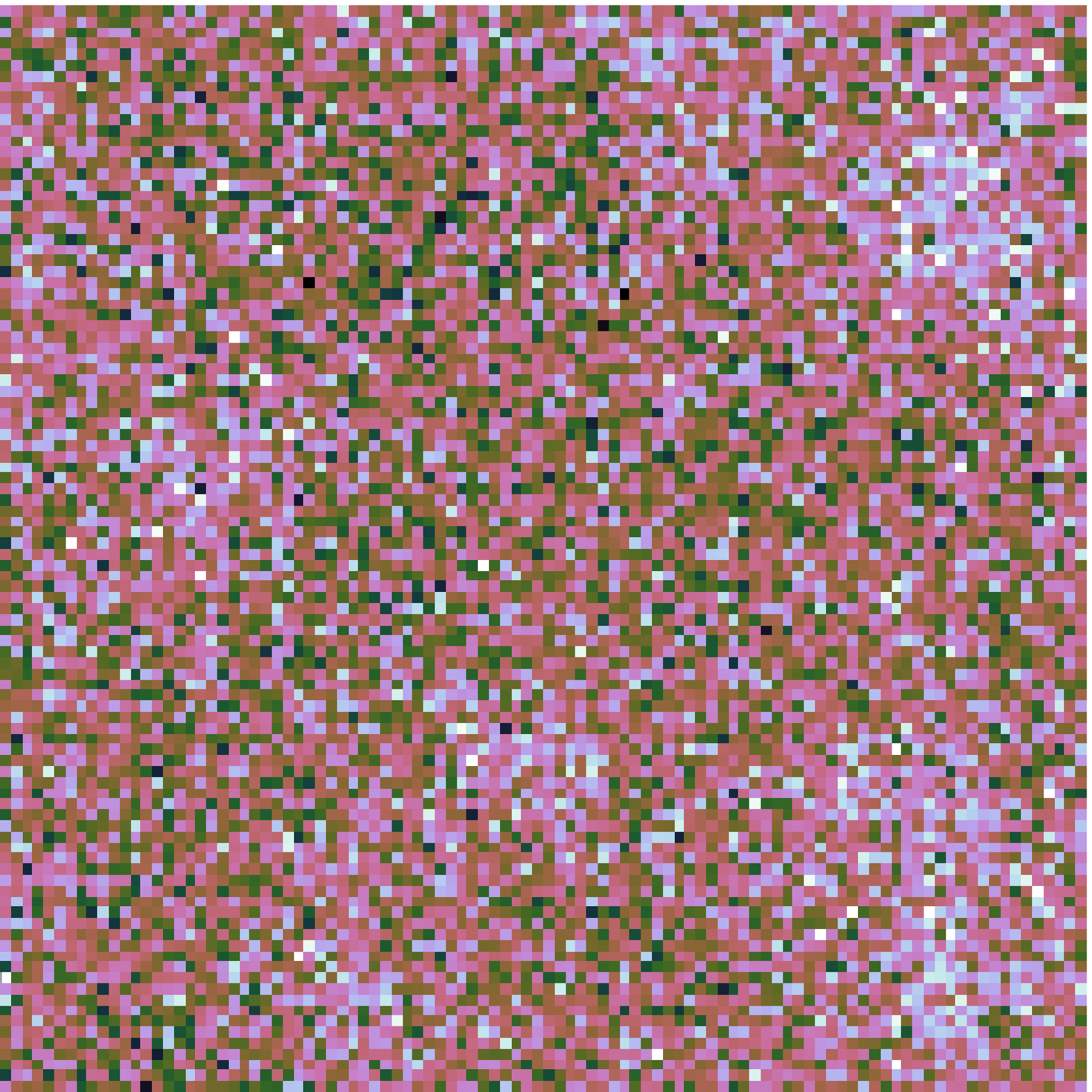} };
	\node (fig3) at (48.1,6)
			{ \includegraphics[trim=0 0 0 0 clip=true,width=0.075\textwidth]{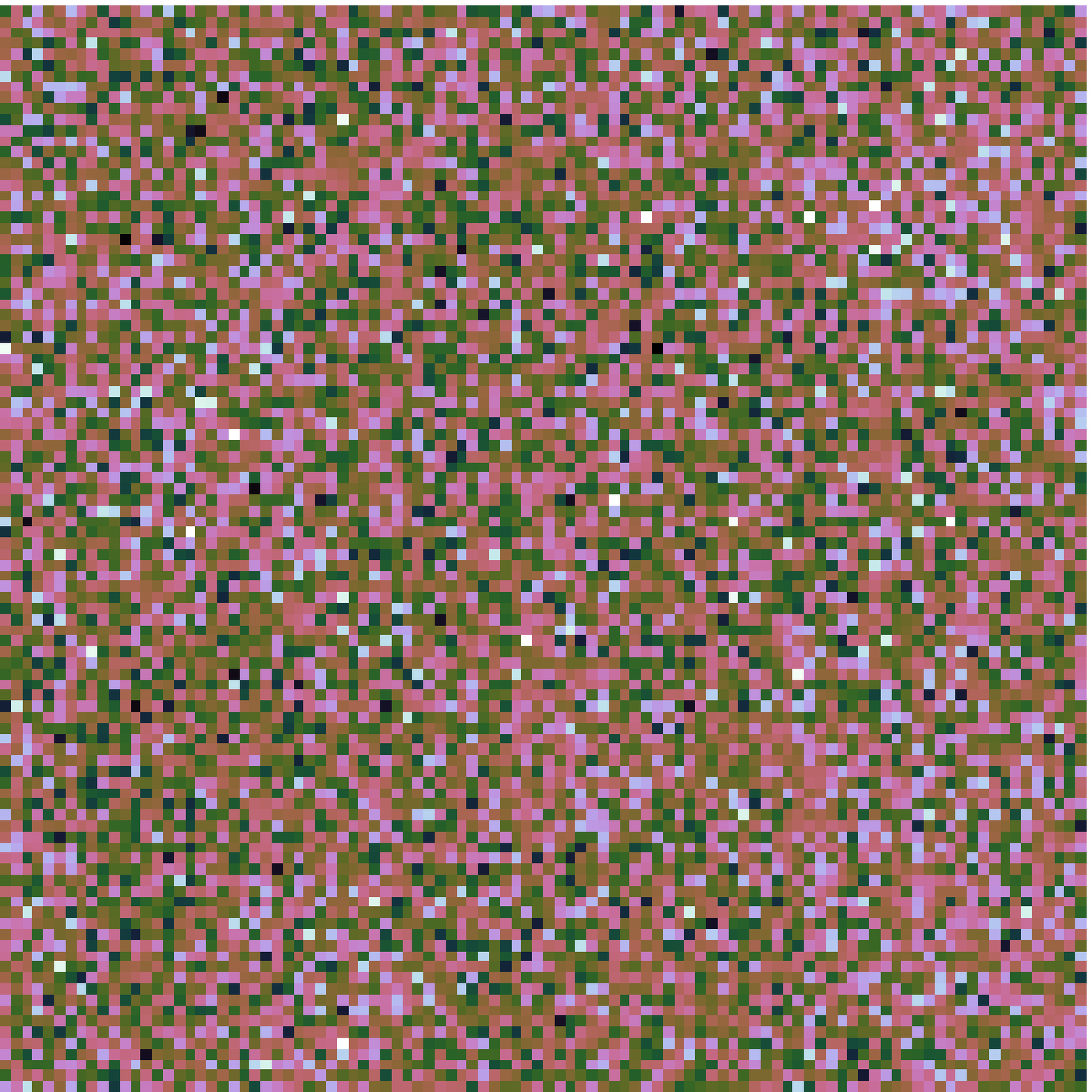} };
		\end{tikzpicture}
\end{tabular}
\caption{The sensitivity of $r_{\rm peak}$ values to noise amplitude, $N/S$ (see Section \ref{subsec:noise}), for the three models. The graphs show mean and 1$\sigma$ of 400 random realizations. The same realizations as in Table \ref{tbl:distribution} with noise amplitude of $N/S=0.1$, $N/S=1$ and $N/S=2$ are shown. As the noise level increases, the accuracy and precision of $r_{\rm peak}$ as an estimator of primary clump size fall off until its measurement becomes dominated by pure noise at $N/S\geq2$.}
\label{fig:modelnoise}
\end{figure*}

 Figure \ref{fig:modelnoise} shows the observable $r_{\rm recovered}$ measured from the noisy images, relative to the observable $r_{\rm peak}$ measured in the same images without noise. The measurements of $r_{\rm recovered}$ are precise, $<7\%$, up to a very high noise amplitude of roughly half the peak flux within primary clumps. In the case where the primary clumps are barely visible ($N/S=1$), the $\mathit{w}$2PF is still able to recover their radii within $18\%$ uncertainty, albeit with a small systematic effect of $3\%$. For much higher noise levels our method fails to accurately recover input clump sizes, but this is expected because the clump structure is completely masked out by Gaussian noise as shown in  bottom right panel of figure \ref{fig:modelnoise}. The DYNAMO-HST maps typically have $N/S < 0.1$ which, as our analysis shows, makes $r_{\rm peak}$ an ideal observable for inferring primary clump size.
 
\begin{figure}[h] 
\centering
\includegraphics[width=0.45\textwidth]{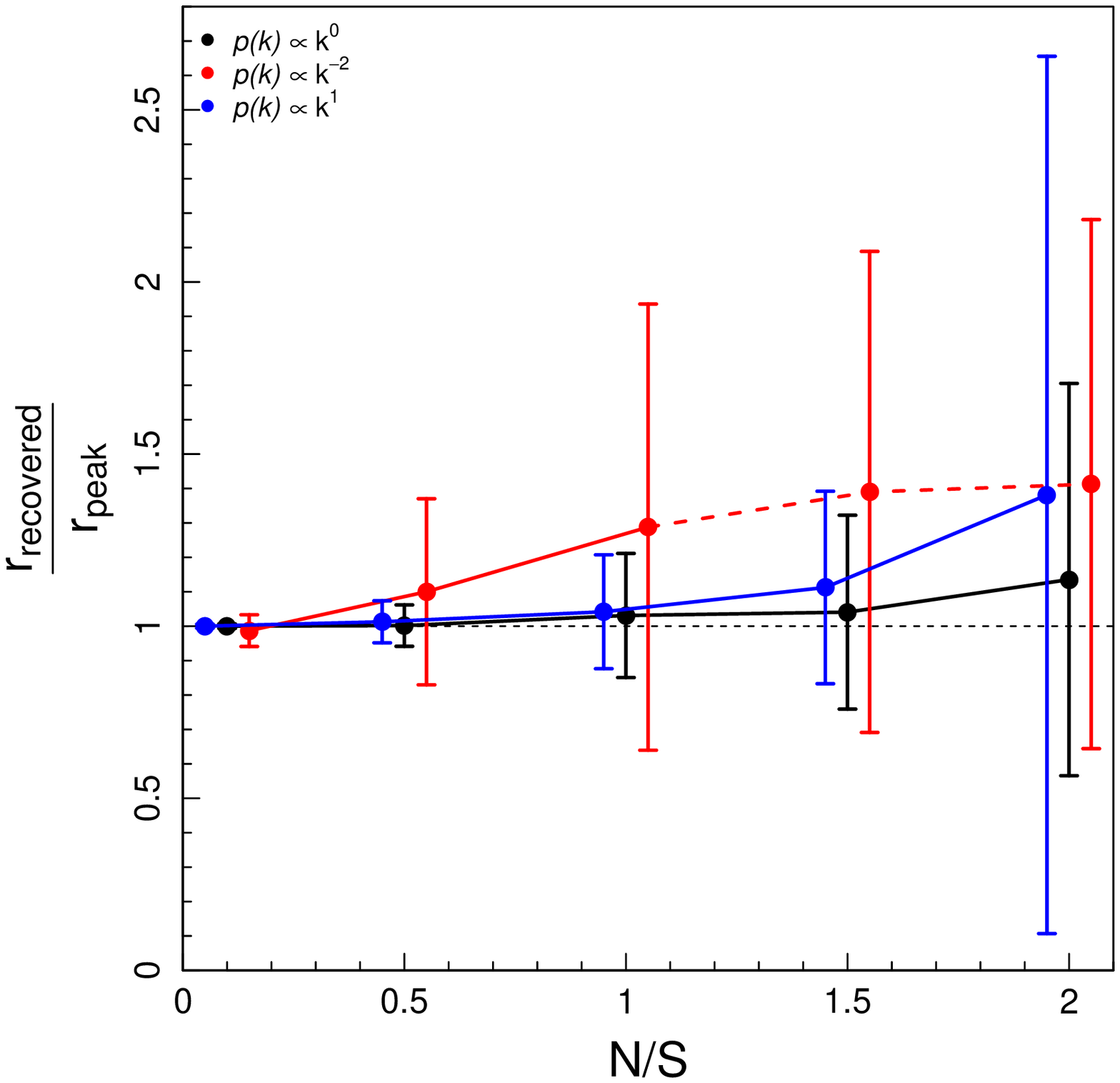}
\caption{The impact of red and blue noise on $r_{\rm peak}$ measurement of the substructure model. The points and error bars indicate mean and 1$\sigma$ of 400 random realization. The 2PCF of pure red noise decays slowly, leading to a monotonically increasing $\mathit{w}$2PF function. Hence, for large noise amplitude ($N/S>1$) our method breaks down because we cannot fit an offset and find the peak value of the raw $\mathit{w}$2PF function indicated by the dashed line.}
\label{fig:rednoise}
\end{figure}

 Since we are using the 2PCF to characterize the clumps, one might wonder about the effect of spatially correlated noise. In optical imaging, pixel noise is normally uncorrelated, i.e. it has a flat power spectrum, but in synthesis imaging the noise has a scale dependence set by the baseline configuration. In Figure \ref{fig:rednoise}, we consider two extreme cases of ``red'' ($p(k)\propto k^{-2}$) and ``blue'' ($p(k)\propto k$) noise. We find that our method remains accurate up to noise levels of $N/S=0.5$ our method remains accurate. Only for strong red noise of $N/S>1$ does the clump scale become seriously masked by this noise.
      
\subsection{Combined Effect of Noise and Blurring} \label{subsec:blur_noise}
   
\begin{figure*} 
\centering
\begin{tabular}{ccc}
	\hspace*{-0.5cm} \includegraphics[trim=0 0 0 0 clip=true,width=0.35\textwidth]{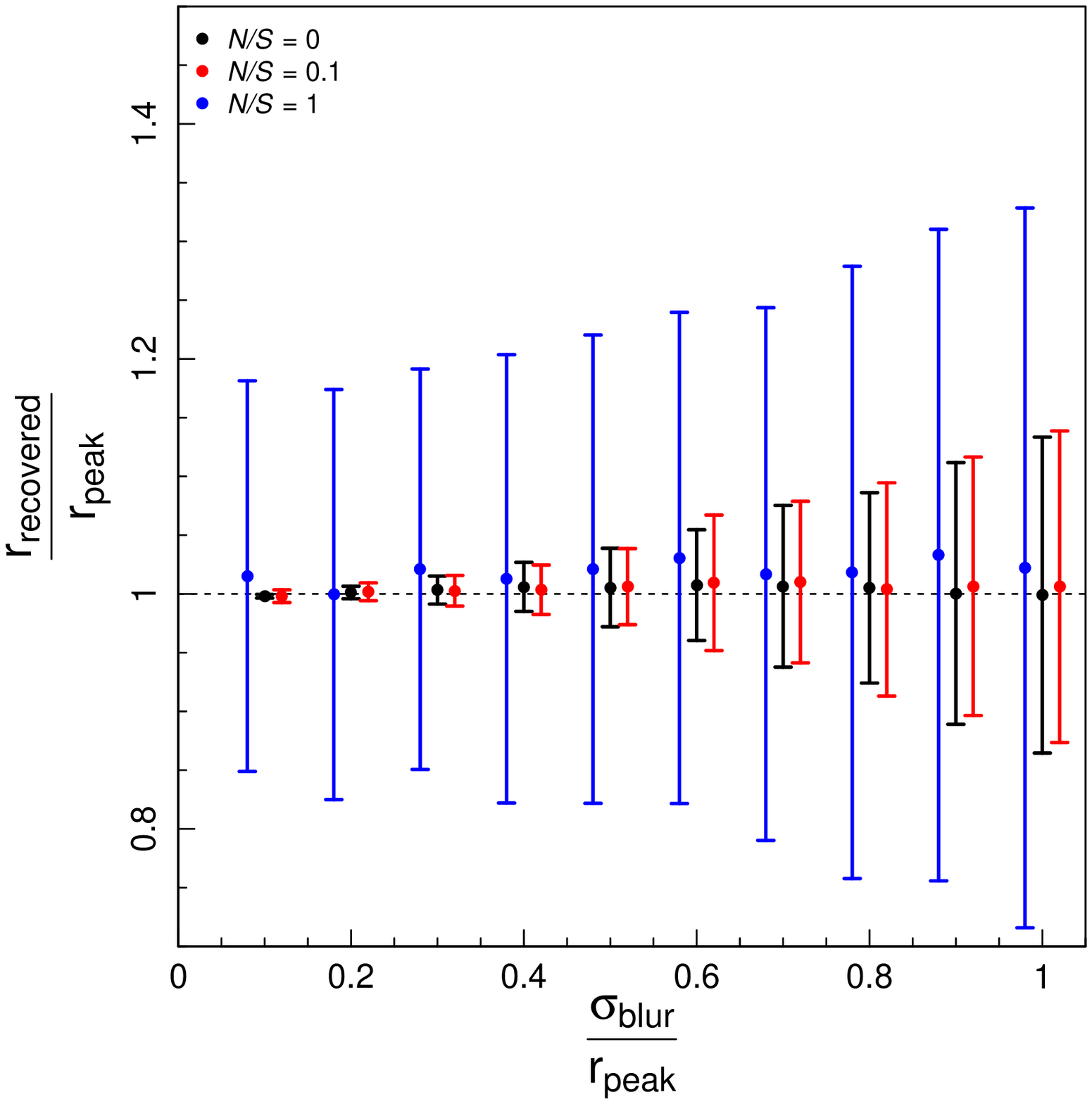}  &
	\hspace*{-0.25cm} \includegraphics[width=0.35\textwidth]{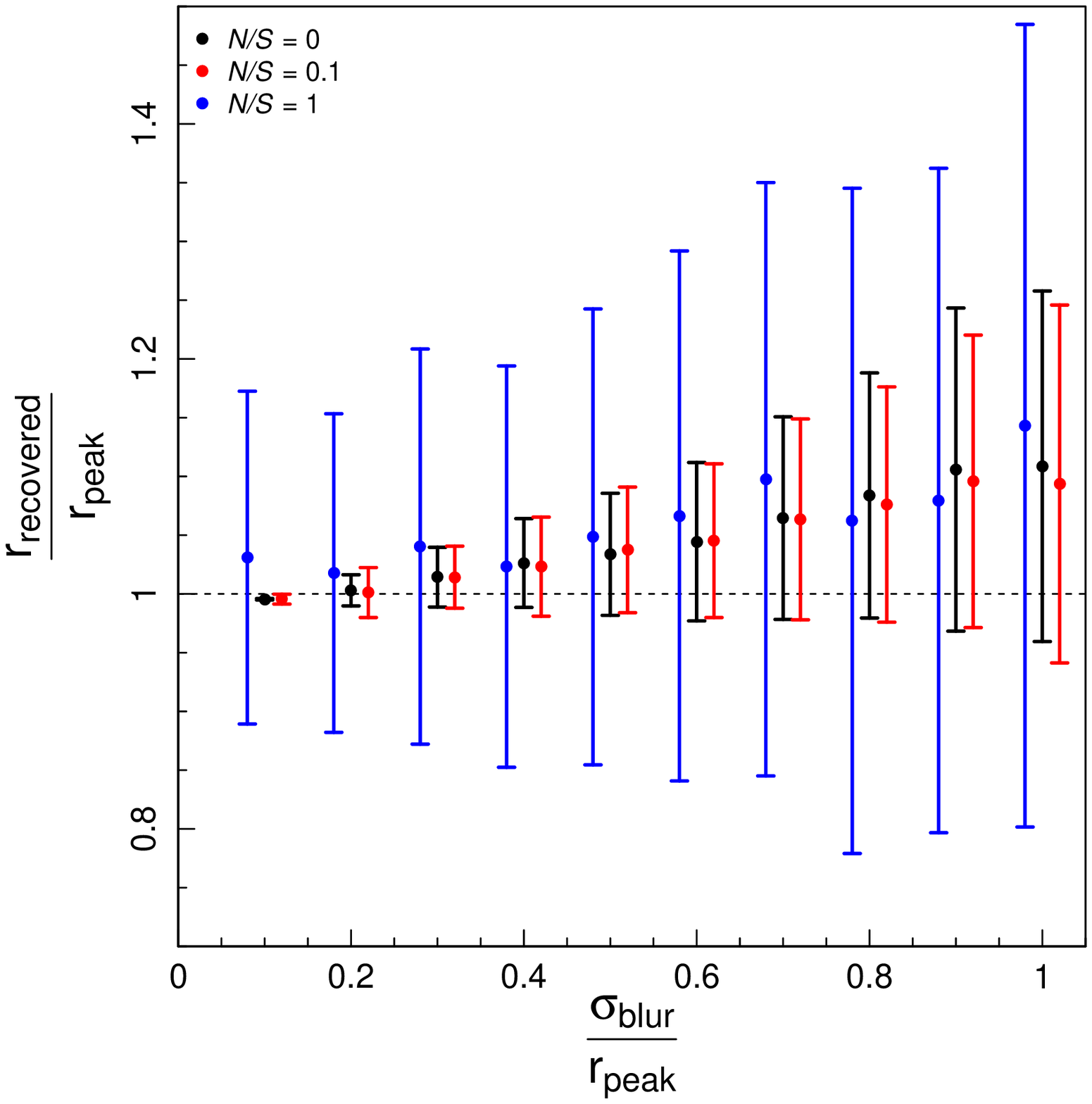} &
	\hspace*{-0.25cm} \includegraphics[width=0.35\textwidth]{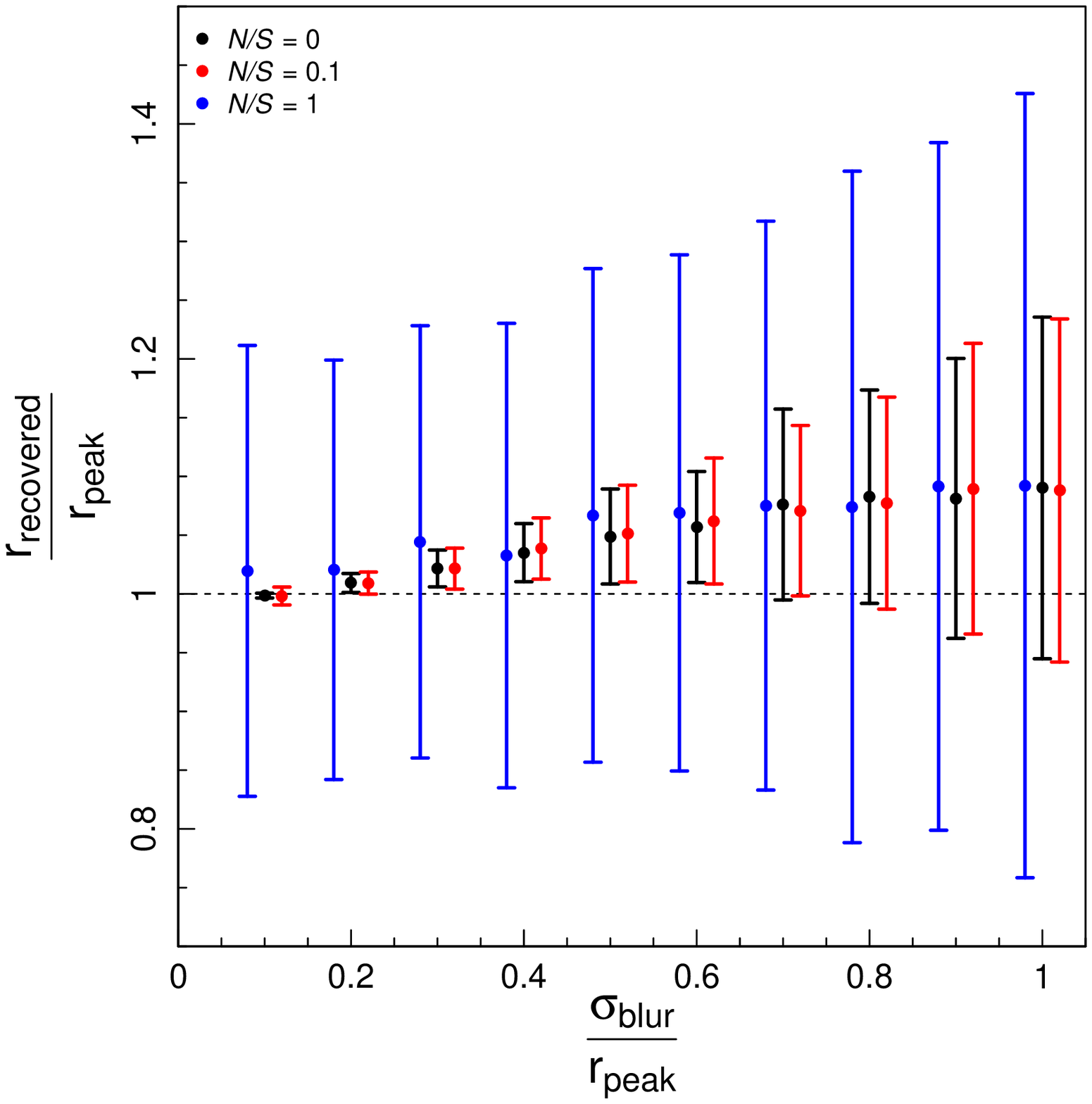} 
\end{tabular}
\caption{The sensitivity of $r_{\rm peak}$ measurement to Gaussian blurring, $\sigma_{\rm blur}$, at different noise amplitudes, $N/S$. The three panels show the three clump models of Table \ref{tbl:distribution}. The points and error bars indicate mean and 1$\sigma$ of 400 random realization.}
\label{fig:modelblur+noise}
\end{figure*}

We now test the hypothesis that the combined effect of noise addition and Gaussian convolution on $r_{\rm peak}$ measurement would be similar to considering their effects independently. Following our earlier definitions, we first apply a blurring scale to each of the three models and then add Gaussian white noise onto the mock $D$-field. The result of our analysis is shown in Figure \ref{fig:modelblur+noise}. 
\\
 In the absence of noise we recover the same behaviour as in Figure \ref{fig:modelblur} for each model, whereas for small blurring scale and large noise amplitude we find the same behaviour as in Figure \ref{fig:modelnoise} (a mean with $\approx 1\%$ systematic offset and a standard deviation of $\approx 18\%$). If we set both sources of error to their maximum considered values, i.e.~$N/S=1$ for Gaussian noise and $\sigma_{\rm blur}/r_{\rm peak}=1$ for blurring, we find a systematic offset dominated by blurring ($\approx 10\%$) and a standard deviation of $\approx 30\%$. The latter is somewhat larger than what is expected from the combination (in quadrature) of the individual noise levels, i.e. $18\%$ for Gaussian noise and $14\%$ for blurring. However, in most realistic scenarios, including the real galaxies analyzed in Sections \ref{sec:M51} and \ref{sec:DYNAMO}, the individual noise levels are low enough ($N/S < 0.1$ and $\sigma_{\rm blur}/r_{\rm peak} \lesssim 0.3$) that their linear combination (in quadrature) can be safely assumed.
 
  \section{Clump size measurements in NGC 5194} \label{sec:M51}
 
 After benchmarking our method of measuring the clump scale using mock density fields, we shall now consider the case of a real galaxy. The aim is to apply the method of Section \ref{sec:LS} to find a typical clump size and compare this measurement to existing measurements based on a clump-by-clump analysis. To this end we chose the main component NGC 5194 of the nearby galaxy system M51, for which detailed H$\alpha$ region analyses are available (\citealp{Guttierrez2011AJ....141..113G}, \citealp{Lee2011ApJ...735...75L}).
   
\subsection{Data}

For our analysis we use the H$\alpha$ image as the data field ($D$) and the F$814$W (\textit{I}-band) image as the normalizing random field ($R$). The H$\alpha$ and continuum maps of NGC 5194 are obtained from the Advanced Camera for Surveys on board the HST (\citealp{Mutchler2005AAS...206.1307M}). We first remove the central bulge of the galaxy, which would otherwise contaminate the analysis due to its strong H$\alpha$ emission. Then we visually remove foreground stars. We also remove the small continuum contamination in the H$\alpha$ image, by subtracting the continuum image from the H$\alpha$ image, ensuring that the H$\alpha$ flux at large radii (beyond the optical disk) falls exponentially to zero. 

From the original HST image we select the region of NGC5194 shown in Figure \ref{fig:m51rdf} (this excludes the companion galaxy NGC5195). The native size of this region is 7000 by 7000 pixels, which we reduce to 2000 by 2000. We do this to reduce the computational time (which scales as the square of the number of pixels). In the image of 2000 by 2000 pixels image each pixel measures $0^{\prime\prime}.17$, which is comfortably smaller than the primary clump size (see Section \ref{subsec:resizeandnoise}), but much larger than the \textit{HST} PSF, hence PSF corrections can be neglected. 
 
 \subsection{Average clump size}
 
\begin{figure*} 
\centering
\begin{tabular}{cc}
	\hspace*{-0.5cm} \raisebox{.105\height}{\includegraphics[trim=0 0 0 0 clip=true,width=0.44\textwidth]{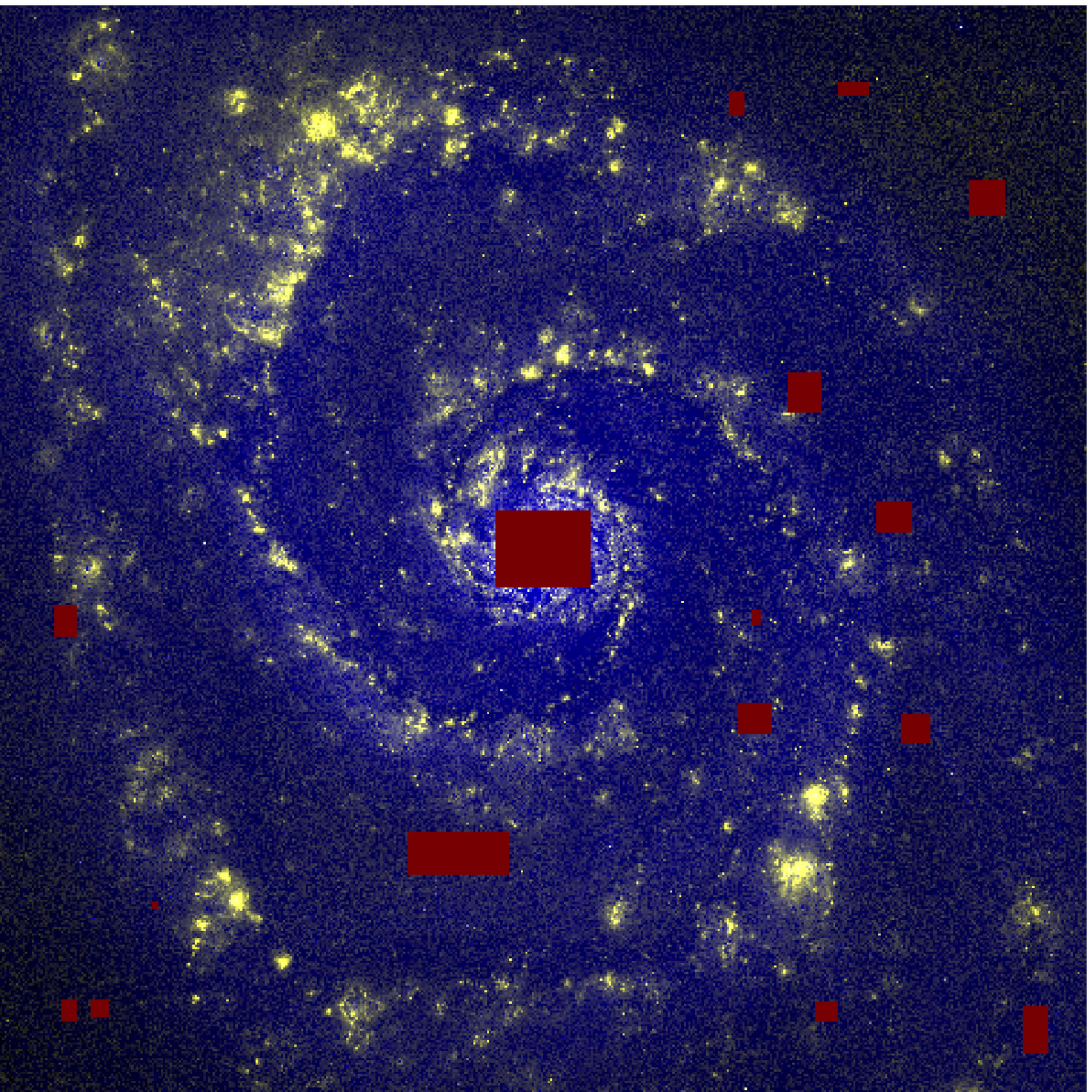} } &
	\includegraphics[width=0.5\textwidth]{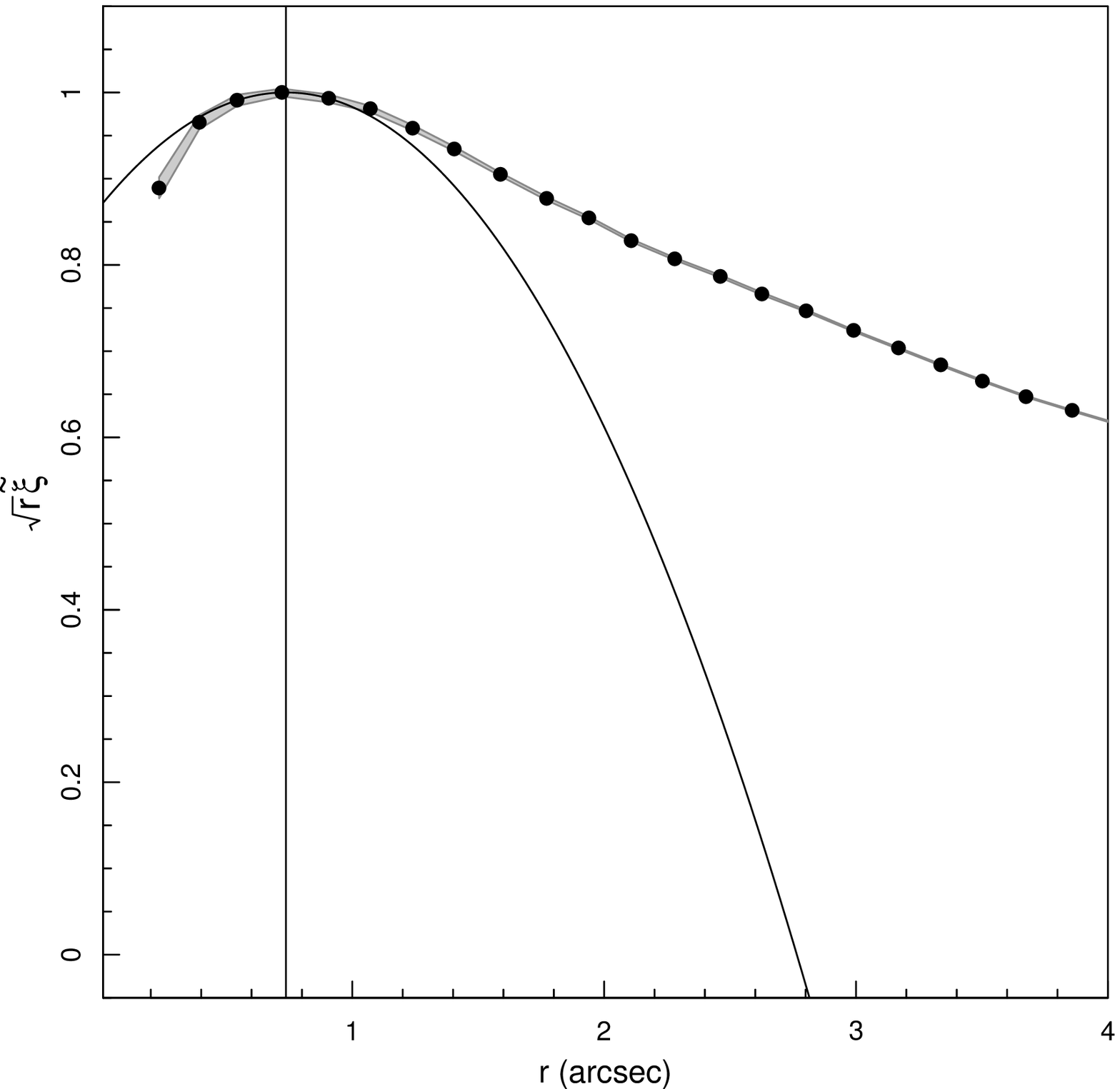} 
\end{tabular}
\caption{Superimposed H$\alpha$ (yellow) and F$814$W continuum (blue) maps of NGC 5194 used in our analysis are presented on the left. The red regions indicate pixels corresponding to bulge and foreground stars masked out before computing the $\mathit{w}$2PF shown on the right. The solid curve and vertical line are the parabolic fit and the inferred $r_{\rm peak}$ value, respectively. The shaded area shows the 1$\sigma$ error on $\mathit{w}$2PF measurement.}
\label{fig:m51rdf}
\end{figure*}

We then follow the method outlined in Section \ref{sec:LS} to recover an average clump size. After computing the $\mathit{w}$2PF, we fit a parabola around global maximum and find $r_{\rm peak} = 0^{\prime\prime}.74\pm0.03$\footnote{We use arcsec in this section because previous NGC 5194 studies use different distance estimates as conversion factors.} as shown in Figure \ref{fig:m51rdf}. The uncertainty in the estimate of $r_{\rm peak}$ is propagated from the uncertainty in the computation of 2PCF.

To compare our estimator with previously measured clump sizes we use the list of radii and luminosities of HII regions measured by \citealp{Guttierrez2011AJ....141..113G}. This study incorporates the circularizing isophotal method whereby the area of a continuous object (connected pixels) with intensity at least three times the rms of the local background is fitted by an equivalent radius, $R_{\rm eq} = \sqrt{A/\pi}$ while the flux within the region is converted into luminosity using a predetermined conversion factor. We compute the $L^2$-weighted average radius of structures less luminous than $10^{38.8} erg\ s^{-1}$, since more luminous ones lie well beyond the break in the clump luminosity function (\citealp{Guttierrez2011AJ....141..113G}) and are normally associated with coincidental agglomerations of uncorrelated clumps. We find an $L^{2}$-weighted radius of $1^{\prime\prime}.99$ with a clump-to-clump standard deviation of $0^{\prime\prime}.91$. We expect this radius to be at least $\sim2$ times larger than $r_{\rm peak}$ due to the method used by Guttierrez: their clump radii are measured by circularizing isophotes, containing almost all the flux in the clumps (without specifying the precise fraction of the luminosity within the isophotes). Assuming that their radii contain $90\%$ of the total clump flux, their radii would be about twice our Gaussian radius, which contains $39\%$ of the total flux.

The analysis of \citealp{Guttierrez2011AJ....141..113G}, shows some disagreement with \citealp{Lee2011ApJ...735...75L}, who find a larger number of clumps and significantly smaller clumps sizes but show that many of the smaller clumps are subclumps. This reinforces the point that conventional clump-by-clump methods measure ever smaller average clump sizes with increasing resolution, while our method recovers a constant size near $\sigma_{\rm max}$ (Section \ref{subsec:SS}), irrespective of the level of substructure that can be resolved.

\subsection{Effect of resizing}  \label{subsec:resizeandnoise}

\begin{figure}[h]
	\centering

		\includegraphics[width=0.45\textwidth]{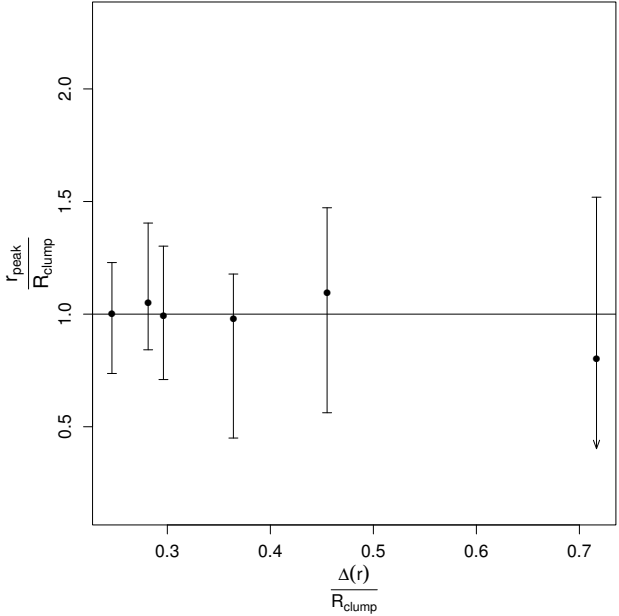} 

\caption{Sensitivity of $r_{\rm peak}$ to the pixel size, $\upDelta(r)$, normalized by $R_{\rm clump}$ ($r_{\rm peak}$ of $0^{\prime\prime}.17$-resolution map). The points indicate the $r_{\rm peak}$ value found using parabolic interpolation. The error bars show the range within which the maximum of $\mathit{w}2PF$ lies (i.e.~independent of the maximum-finding algorithm). Resizing the NGC 5194 map only increases the variance when measuring $r_{\rm peak}$ as seen from the lack of a systematic shift in the allowable range for $r_{\rm peak}$ at different bin widths.}
\label{fig:m51resoandnoise}
\end{figure}

As a sanity check, we wish to quantify the sensitivity of the estimator $r_{\rm peak}$ to resizing of the galaxy map (to less than 2000 by 2000 pixels). Will decreasing the number of pixels lead to measuring a larger value of $r_{\rm peak}$? We test this by defining $r_{\rm peak}$ measured from the 2000 by 2000 image ($0^{\prime\prime}.17$ resolution) as the reference value denoted by $R_{\rm clump}$. We then reduce the size of said maps into lower resolution images and compare the recovered value of $r_{\rm peak}$ with $R_{\rm clump}$ as shown in Figure \ref{fig:m51resoandnoise}. It is apparent that the global maximum of the $\mathit{w}$2PF falls within the same range even for the case where resolution is similar to $R_{\rm clump}$. Hence computing the $\mathit{w}$2PF should give the same result, irrespective of the level of substructure resolved within the clumps. (Of course, at least the primary clumps should be roughly resolved.)

\section{Clump size measurements in DYNAMO-HST Galaxies} \label{sec:DYNAMO}

Having tested its robustness, we now use our method to measure the mean clump sizes in three very clumpy galaxies, drawn from the DYNAMO-\textit{HST} sample and shown in Figure~\ref{fig:Dynamo} . The galaxies in question were observed on the HST Advanced Camera for Surveys Wide-field Camera using the ramp filters FR716N and FR782N to target H$\alpha$ emission within a 2\% bandwidth. The associated FR647M filter was used to generate a continuum image and subtract it from the H$\alpha$ map. The integration times for the H$\alpha$ and continuum images were 45 minutes and 15 minutes, respectively. The full reduction and analysis of the observed data are presented in \citealp{Fisher2017MNRAS.464..491F}. 

These galaxies are local analogs of main-sequence star-forming galaxies of redshift $z\approx1.5$ with rotating, disk-like kinematics. The advantage of using these galaxies over high-$z$ disks lies in their proximity. The adaptive optics observations of H$\alpha$ typically achieve an FWHM resolution of $0^{\prime\prime}.15-0^{\prime\prime}.2$ (e.g \citealp{Wisnioski2012}, \citealp{Genzel2011ApJ...733..101G} ). This corresponds to an image with Gaussian PSF of standard deviation $500-700~pc$ at $z=1.5$. In comparison the standard deviation of the DYNAMO-HST sample is $60-130~pc$, about a tenfold increase in resolution. We assume that the clump-by-clump analysis of \citealp{Fisher2017ApJ...839L...5F} contains most, if not all, of the primary clumps, the output of which can be compared to the estimator $r_{\rm peak}$.

As in the case of NGC 5194 we apply the procedure developed in Section \ref{sec:LS} to the HST maps of the galaxies D13-5, G04-1, and G20-2. After measuring the value of $r_{\rm peak}$ we need to remove the contribution from the PSF, which was insignificant for the \textit{HST} map of NGC 5194. We assume a Gaussian PSF of standard deviation $0^{\prime\prime}.037$, matching the observed FWHM of $0^{\prime\prime}.088$ (\citealp{Fisher2017MNRAS.464..491F}), and adjust $r_{\rm peak}$ by subtracting this value in quadrature -- about a $20\%$ correction. The final estimates of $r_{\rm peak}$ are given in Table \ref{table:params} along with the mean and the standard deviation of the clump radii measured by \citealp{Fisher2017MNRAS.464..491F}. Their technique involves identifying peaks at least three times larger than a smoothed mask as clumpy structure. These regions are then fit iteratively by a 2D Gaussian with a baseline beyond four times that of the major axis of the ellipse. Since the assumed flux profiles are Gaussian this allows us to compare our raw $r_{\rm peak}$ measurements with those of \citealp{Fisher2017MNRAS.464..491F}.

\begin{table}[H]
\def\arraystretch{2}
\centering
\begin{tabular}{|C{1.5cm}|C{2.5cm} | C{2.5cm} |}
\hline
 Galaxy & $r_{\rm peak}$ (pc) & $\frac{1}{2} \bar{D}_{\rm core}$ (pc)\\ \hline
 \begin{tabular}{@{}c@{}} D13-5  \\  D13-5*  \end{tabular} & \begin{tabular}{@{}c@{}} $319^{+64}_{\rm -67}$  \\ $235^{+47}_{\rm -50}$  \end{tabular}  & $206 \pm 72$ \\  \hline
 G04-1 & $295^{+59}_{\rm -66} $ & $236 \pm 118$ \\ \hline
 G20-2 & $307^{+62}_{\rm -68} $ & $329 \pm 148$\\ \hline
\end{tabular}
\caption{Average clump radii of three DYNAMO-HST galaxies, measured by our method ($r_{\rm peak}$) and Mean of the clump-by-clump measurements of \citealp{Fisher2017MNRAS.464..491F} ($\frac{1}{2} \bar{D}_{\rm core}$).}
\label{table:params}
\end{table}

The uncertainty ranges are 68\% confidence intervals accounting for (1) sample variance, (2) deblurring errors, and (3) image noise. \textit{Sample variance} refers to the fact that each measurement is based on only one galaxy with a finite number of clumps, i.e. on one instance of a statistical ensemble. The uncertainty due to the scatter of this ensemble (about $20\%$), assumed to be the same for each galaxy, is taken from our numerical analysis in Table \ref{tbl:distribution} (bottom right panel). By construction, this sample variance includes fitting errors of $r_{\rm peak}$ in the $\mathit{w}$2PF. \textit{Deblurring errors} refer to the uncertainty introduced when correcting for the PSF. These errors (about $\pm1\%$, with a systematic component of $+1\%$) are taken from the numerical analysis in Figure \ref{fig:modelblur} (right panel) for each galaxy. Finally, \textit{image noise} is the uncertainty due to noise in the H$\alpha$ map. Using the definition given in Section \ref{subsec:noise}, we find noise amplitudes of $0.03,\ 0.06$ and $0.075$ for galaxies D13-5, G04-1 and G20-2, respectively, which correspond to errors of $<0.5\%$ and negligible compared to other sources.

By design of our method, large coherent H$\alpha$ structures not reflected in the stellar continuum affect the clump size measurement. This is apparent in two of our galaxies. First, D13-5 shows a bright chain of H$\alpha$ clumps stretching into the third (bottom left) quadrant. This quadrant contains a 50\% excess flux relative to the other quadrants, which is not reflected in the continuum map. Removing this quadrant from the analysis steepens the $\mathit{w}$2PF significantly (red line in Figure~\ref{fig:Dynamo}) and decreases the clump size by about $27\%$. This is the case labeled D13-5* in Table \ref{table:params}. Removing any other quadrant has only an insignificant effect. Second, the galaxy G04-1 shows an extended (1-2 kpc) `sea' of H$\alpha$ in the first (top right) quadrant, which is also not seen in the continuum. This feature causes the plateauing of the $\mathit{w}$2PF with a weak secondary maximum around 1.2-1.4 kpc. Unlike in the previous case, removing this region from the analysis has no significant effect on the clump size measurement, i.e.~on the position of the absolute maximum of the $\mathit{w}$2PF. We conclude that if the $\mathit{w}$2PF is relatively flat (i.e.~it changes by less than $\sim10\%$ from $r_{\rm peak}$ to $2r_{\rm peak}$), it is advisable to check whether any large structures in the $D$-field not seen in the $R$-field have affected the measurement and consider removing/modeling them.

Comparison between our measurements of $r_{\rm peak}$ and the clump sizes presented in \citealp{Fisher2017ApJ...839L...5F} shows good agreement between the two sets of values (Table \ref{table:params} and Figure~\ref{fig:Dynamo}). It should be noted that the comparison values from \citealp{Fisher2017ApJ...839L...5F} (right in Table \ref{table:params}) are arithmetic means of the clump radii rather than luminosity-weighted averages. This is justified, because at the present resolution only the primary clump generation can be resolved and not its substructure. If the resolution were increased to resolve substructure, the mean size would drop, whereas our method would still recover the same value (within statistical uncertainties).

\newcommand{\imagepastetwo}[1]{ \hspace*{-0.5cm} \includegraphics[trim=0 0 0 10cm, clip=true,width=0.35\textwidth]{#1} } 
\newcommand{\imagepastethree}[1]{ \hspace*{-0.5cm} \includegraphics[trim=0 0 0 10cm, clip=true,width=0.36\textwidth]{#1} } 
\newcommand{\graphpastetwo}[1]{ \hspace*{-0.5cm} \includegraphics[trim=0 0 0 0cm, clip=true,width=0.35\textwidth]{#1} }

\begin{figure*}
\centering
\begin{tabular}{ccc}
	D13-5 & G04-1 & G20-2 \\
	\imagepastetwo{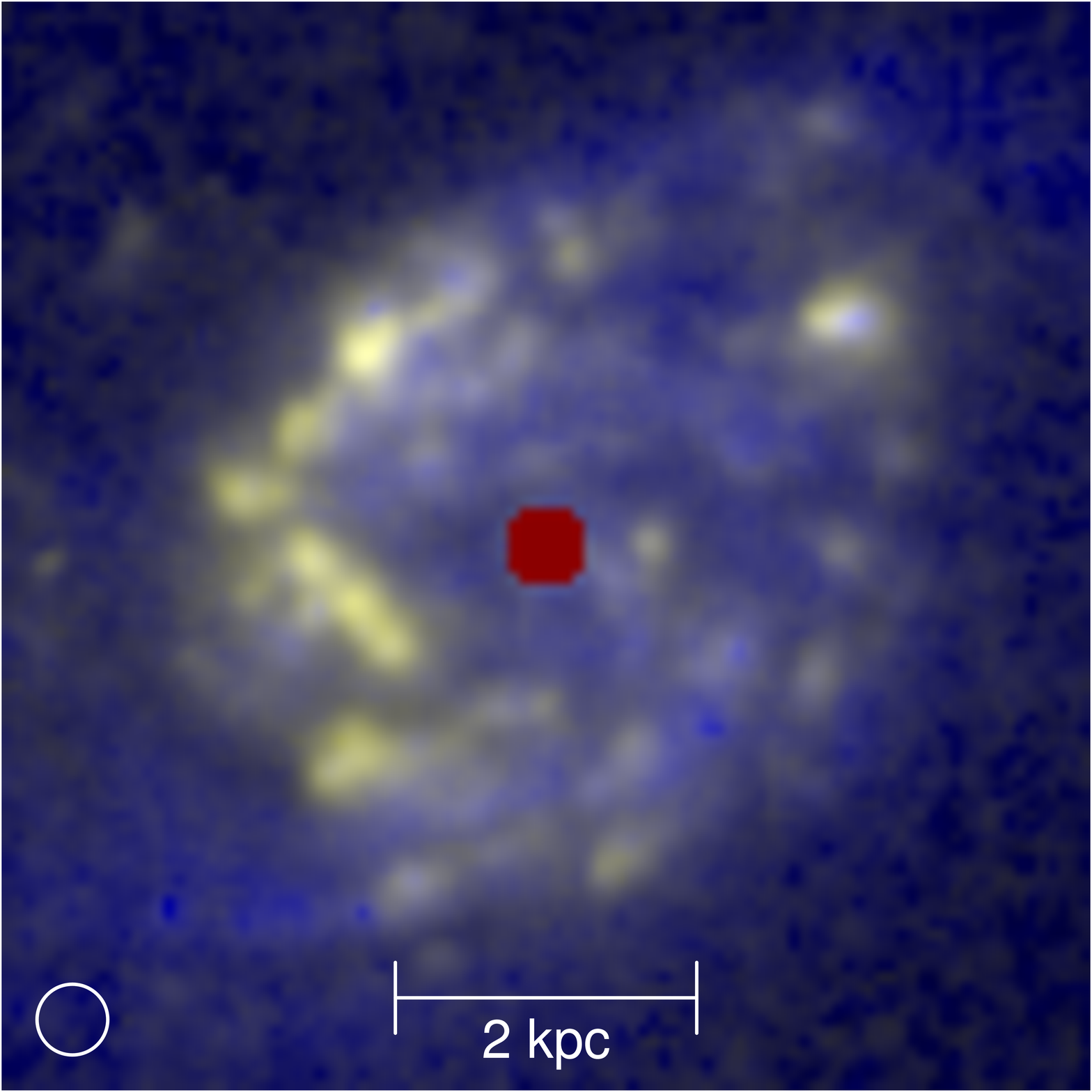} & \imagepastetwo{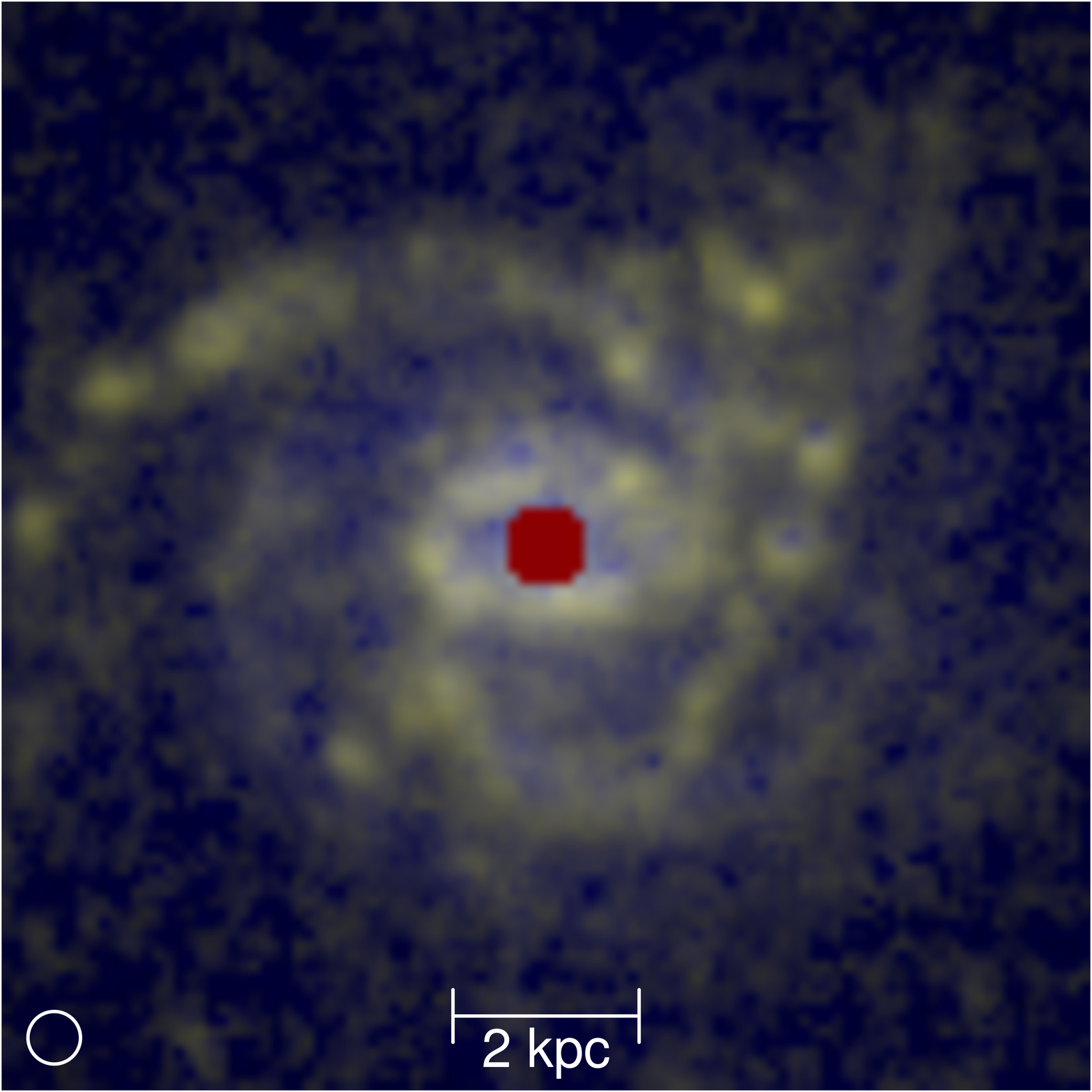} & \imagepastethree{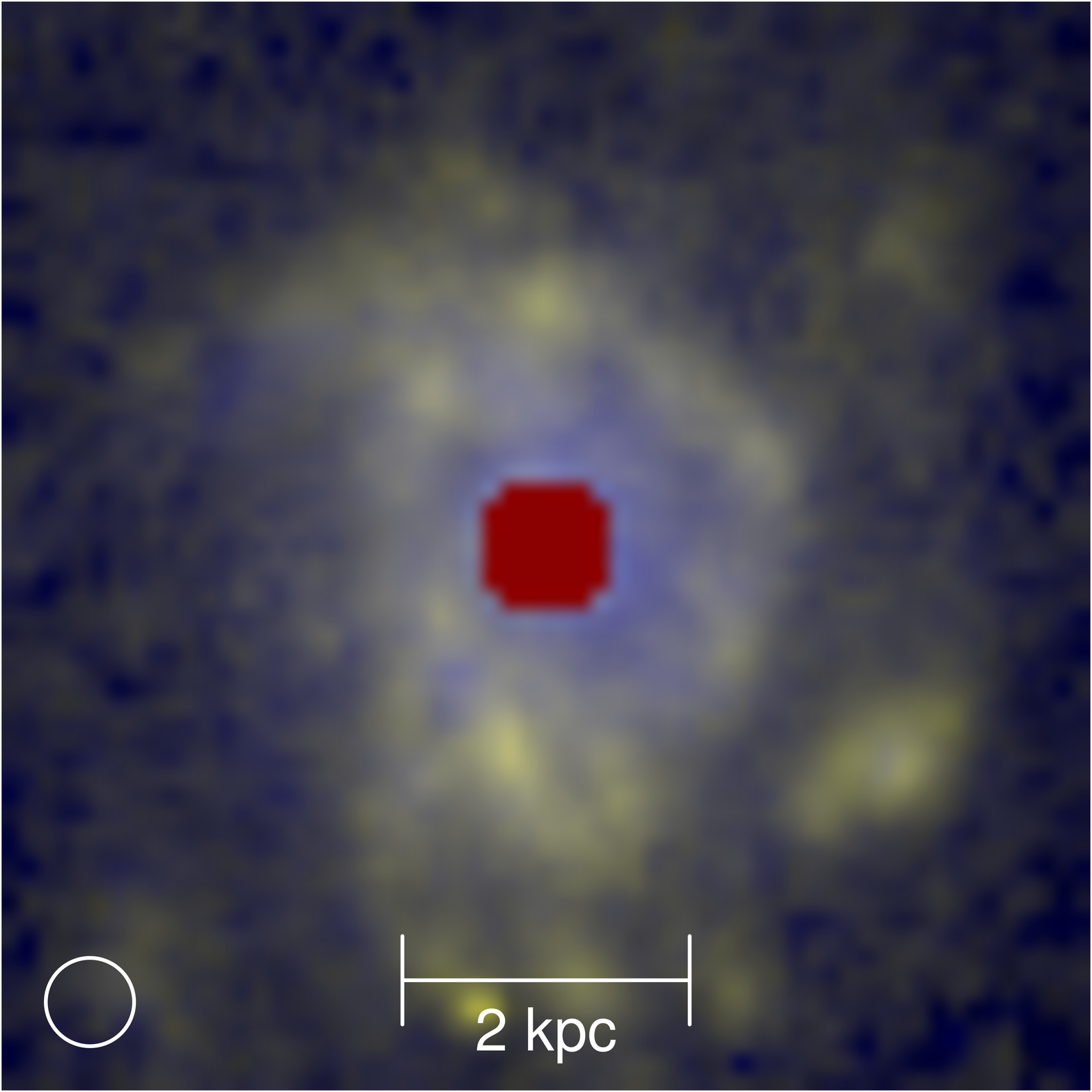} \\
	\graphpastetwo{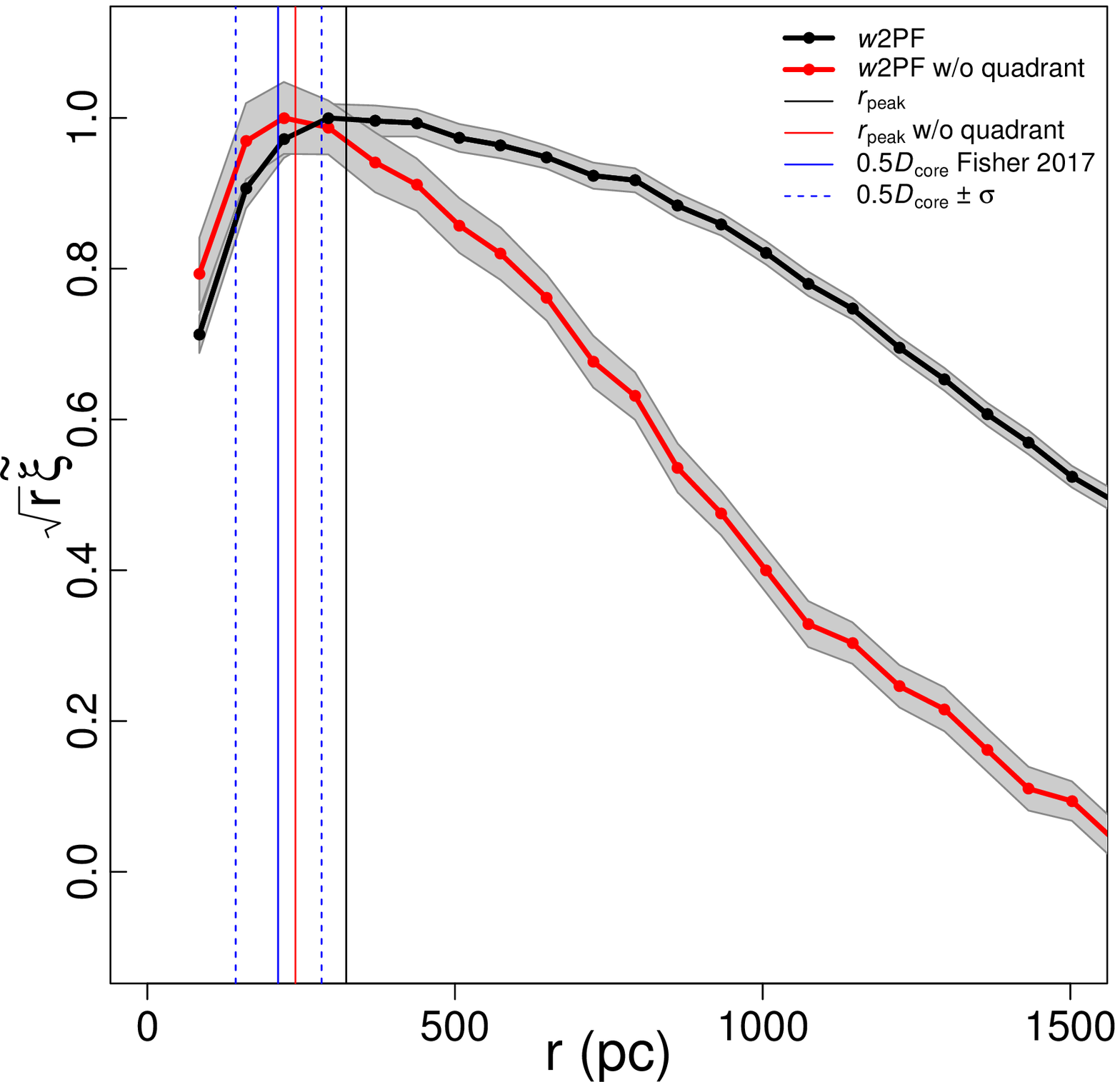} & \graphpastetwo{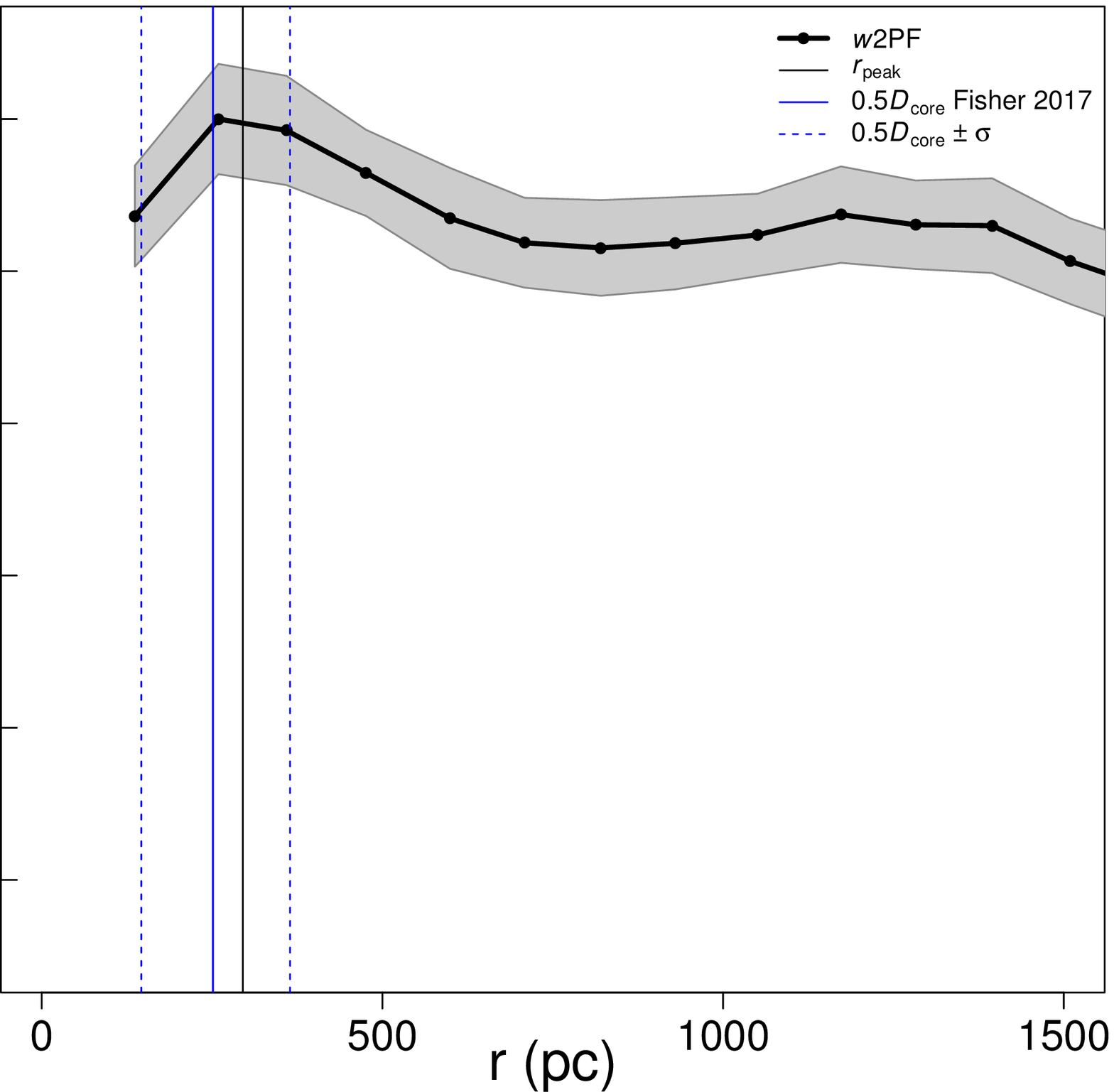} & \graphpastetwo{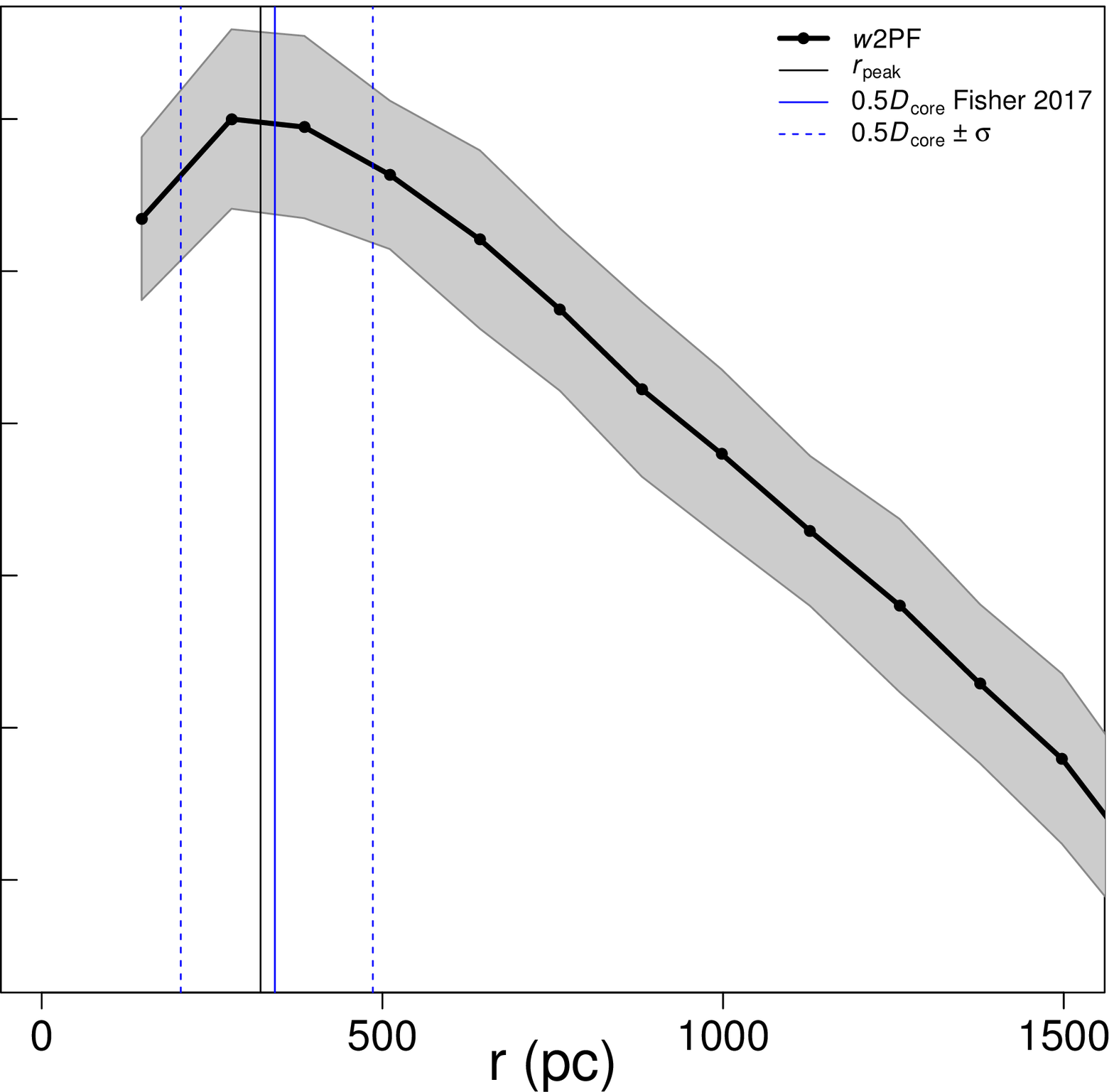}
\end{tabular}
\caption{Top row: HST H$\alpha$ (yellow) map superimposed with FR$647$M continuum image (blue) and the bulge masked out (dark red) for the three DYNAMO galaxies analyzed here. Bottom row: their respective $\mathit{w}$2PF. Similarly to NGC 5194, the $\mathit{w}$2PF (connected black dots) is fitted with a parabola (not shown) to find the peak position $r_{\rm peak}$ (black vertical line), which is interpreted as the primary clump radius. The shaded region represents 1$\sigma$ deviation on $\mathit{w}$2PF measurement, calculated directly from the pair counts (DD, DR, RR) in the 2PCF. Note that these errors are strongly correlated between different values of $r$, hence we do not use these error bars directly for the uncertainty calculations (see text). On the galaxy images, the white circles have a radius $r_{\rm peak}$, corrected for the PSF. The mean size of the primary clumps (i.e.~without substructure) measured on a clump-by-clump basis (\citealp{Fisher2017MNRAS.464..491F}) is shown as blue vertical line, with dotted lines indicating the 1$\sigma$ scatter. The H$\alpha$ field of galaxy D13-5 shows a strong excess in the bottom left quadrant (also relative to the continuum map). Removing this quadrant produces the $\mathit{w}$2PF shown as the red curve and the reduced $r_{\rm peak}$ value represented by the red vertical line.}
\label{fig:Dynamo}
\end{figure*}

\section{Discussion and Conclusions}

In this paper we have applied the two-point statistics, more commonly used in cosmology, to measure the scale of star-forming regions in galaxy images, specifically H$\alpha$ emission maps. The main challenge in this approach is that star formation maps also contain structure, such as spiral arms and the exponential profile of the disk. This situation is analogous to measuring the two-point statistics in cosmology in the presence of a galaxy selection function. We therefore import the cosmological solution to this problem and use the map of an old stellar population as normalizing random field, which serves as a baseline for the two-point statistics.

We found that the radius-weighted two-point correlation function $\mathit{w}$2PF is well suited to recover the primary clump scale (i.e.~without contamination from substructure). The method recovers this primary clump scale irrespective of how much substructure is resolved. In particular, this means that our method enables a robust comparison of samples at different redshifts, and it enables for a direct comparison between the primary clump scale and global instability scales (Jeans and shear lengths). These are significant advantages over traditional, more subjective methods. An additional strength of our method is its robustness against noise: even noise levels that make individual clumps difficult to identify  still allow for a statistical recovery of the primary clump scale.

On the downside, the two-point statistics does not allow us to analyse the individual clumps, but only their global statistics. Hence this method is particularly suitable for large samples of galaxies, e.g.~covering a range of redshifts and/or masses, to analyse galaxy-to-galaxy variations in clump sizes. Such a sample could exploit other indicators of star formation than H$\alpha$, for instance UV, radio, and CO emission, as well as to other tracers of stellar density than FR$647$M. If no suitable stellar map is available, one could resort to using a disk model (e.g.~an exponential profile) as the normalizing field $R$. We ran a few tests of this idea, indicating that this a promising avenue that we plan to explore in greater detail as we need to take into account the asymmetric clump distribution observed in clumpy galaxies. In a forthcoming paper we will explore this road using a large sample of clumpy galaxies with different physical conditions.

\acknowledgments{
  K.A was supported by the Research Collaboration Award 12105205 of the University of Western Australia. R.G.A thanks NSERC and the Dunlap Institute for financial support. D.B.F and K.G acknowledge support from Australian Research Council grants DP130101460 and DP160102235. The Hubble Space Telescope data in this program are drawn from the HST program PID 12977 (PI Damjanov). We thank our referee, Erik Rosolowsky, for insightful comments.}

\clearpage

\begin{appendix}
\section{Derivation of the expectations} \label{app:expectation}

We start by defining the Fourier transform (FT) and its inverse (IFFT) similar to \citealp{Peacock1999coph.book.....P} albeit with a change in sign of $i$.
\begin{equation}
\mathcal{F}(\pmb{k}) = \frac{1}{V} \int \mathcal{F}(\pmb{r})\ e^{-i\pmb{k}\cdot \pmb{r}}d^Dr 
\end{equation}
\begin{equation}
\mathcal{F}(\pmb{r}) = \frac{V}{(2\pi)^D} \int \mathcal{F}(\pmb{k})\ e^{i\pmb{k}\cdot \pmb{r}}d^Dr
\end{equation}
where $\pmb{r},\pmb{k} \in \mathbb{R}^D$ are the real-space and wave vectors, respectively, and $V$ is the volume of the real-space domain. Now let us consider a function, $\mathcal{F}(\pmb{r})$, composed of superposition of multiple fields defined on the same domain. We can explicitly write such a function as
\begin{equation}
\mathcal{F}(\pmb{r})=\sum_{\rm l=1}^{N}\delta_l(\pmb{r}),
\end{equation}
where $N$ and $\delta$ represent the total number and functional form of the fields, respectively. The power spectrum of such a function is given by
\begin{equation} \label{appeq:ps}
\left\langle P(\pmb{k})\right\rangle = \left\langle\mathcal{F}(\pmb{k})\mathcal{F}^{\dagger}(\pmb{k})\right\rangle = \frac{1}{V^2}\left( \sum_{\rm l=1}^{N}\left\langle|\delta_l(\pmb{k})|^2\right\rangle + \sum_{\rm l=1}^{N}\sum_{\rm \substack{m=1 \\ m\neq l}}^{N}\left\langle\delta_l(\pmb{k})\delta^{\dagger}_m(\pmb{k})\right\rangle\right) 
\end{equation}
where we have used the involutory property of conjugates and linearity of expectation to separate out the terms representing power and cross spectra. This expression allows us to work with the profile of individual fields to get an idea of the overall 2PCF. 

\section{Renormalized 2PCF} \label{app:renormal}

The mean zero-density field of a Gaussian clump with size parameter $\sigma$ can be described by
\begin{equation}
\delta_\sigma(\pmb{r}) = \frac{V}{2\pi\sigma^2}e^{-\frac{(\pmb{r}-\pmb{\mu})^2}{2\sigma^2}} - V,
\end{equation}
Given the fact that the FTs of a Gaussian and a constant are a Gaussian and a Dirac delta function, respectively (\citealp{bracewell1978fourier}), we can infer that the power spectrum would also be a Gaussian. This leads to an isotropic 2PCF of
\begin{equation}
\xi_\sigma(r) = V\left(\frac{1}{2\pi\left(\sqrt{2}\sigma\right)^2} e^{-\frac{r^2}{2\left(\sqrt{2}\sigma\right)^2}} - 1\right)
\end{equation}
for a Gaussian clump defined with mean zero. Hence, we recover a 2PCF stretched by a factor of $\sqrt{2}$ with an offset that depends on the normalizing scheme of the initial density field. However, in order to extract the correct clump size from the $\mathit{w}$2PF we would like to remove this offset.

The renormalisation step is simple in the case of a single Gaussian clump where adding unity to the 2PCF, after removing the volume contribution, gets rid of the offset. However, is the amplitude of this offset the same in the case of multiple Gaussian clumps? We check this by defining a the density field of a model consisting of multiple Gaussians as
\begin{equation}
\mathcal{F}(\pmb{r})=\frac{1}{N} \sum_{\rm l=1}^{N}\delta_{\rm \sigma_{\rm l}}(\pmb{r}) - 1.
\end{equation}
We again use Eq.~(\ref{appeq:ps}) to find the power spectrum of this model
\begin{equation}
\begin{split}
P(\pmb{k})  & = \left(\frac{1}{N}\sum_{\rm l=1}^{N} \delta_{\rm \sigma_{\rm l}}(\pmb{k}) - \updelta_{D}(\pmb{k})\right)\left(\frac{1}{N}\sum_{\rm l=1}^{N}\delta_{\rm \sigma_{\rm l}}(\pmb{k}) - \updelta_{D}(\pmb{k})\right)^{\dagger} \\
& = \frac{1}{N^2}\sum_{\rm l=1}^{N}|\delta_{\rm \sigma_{\rm l}}(\pmb{k})|^2 + \frac{1}{N^2}\sum_{\rm l=1}^{N}\sum_{\rm {\substack{m=1 \\ m\neq l}}}^{N}\delta_{\rm \sigma_{\rm l}}(\pmb{k})\delta^\dagger_{\rm \sigma_{\rm m}}(\pmb{k}) - \updelta_{D}(\pmb{k})\frac{1}{N}\sum_{\rm l=1}^{N}\delta^\dagger_{\rm \sigma_{\rm l}}(\pmb{k}) - \frac{1}{N}\sum_{\rm l=1}^{N}\delta_{\rm \sigma_{\rm l}}(\pmb{k}) \updelta_{D}^\dagger(\pmb{k}) + \updelta_{D}(\pmb{k})\updelta_{D}(\pmb{k})^{\dagger}
\end{split}
\end{equation}
The first and second terms are the summation of individual power and cross spectra, respectively, while the subsequent terms can be written in terms of a Dirac function. With a little algebra we can simplify the cross-correlation terms to the expression
\begin{equation}
\begin{split}
\frac{1}{N^2}\sum_{\rm l=1}^{N}\sum_{\rm {\substack{m=1 \\ m\neq l}}}^{N}\delta_{\rm \sigma_{\rm l}}(\pmb{k})\delta^\dagger_{\rm \sigma_{\rm m}}(\pmb{k}) & = \frac{2}{N^2}\left(\sum_{\rm l=1}^{N-1}\sum_{\rm m=l+1}^{N} e^{-\frac{k^2(\sigma^2_l+\sigma^2_m)}{2}}\left\langle\cos(\pmb{k}\cdot(\pmb{\mu}_l-\pmb{\mu}_m))\right\rangle\right). \\
\end{split}
\end{equation}
Under the assumption of randomly distributed clump centers the expectation of the cosine term vanishes for $\pmb{k}\neq0$. For this uncorrelated model, we can further simplify the cross-correlation part of the equation in terms of a Dirac delta function:
\begin{equation}
\begin{split}
\frac{1}{N^2}\sum_{\rm l=1}^{N}\sum_{\rm {\substack{m=1 \\ m\neq l}}}^{N}\delta_{\rm \sigma_{\rm l}}(\pmb{k})\delta^\dagger_{\rm \sigma_{\rm m}}(\pmb{k}) & = \updelta_{D}(\pmb{k})\frac{2}{N^2}\left(\frac{N(N-1)}{2}\right) \\
\end{split}
\end{equation}
Collecting all the terms involving $\updelta_{D}(\pmb{k})$ we can write the resulting isotropic power spectrum of the uncorrelated multiple Gaussian model as
\begin{equation}
p(k)  = \frac{1}{N^2} \sum_{\rm l=1}^{N} p_{\rm \sigma_{\rm l}}(k) - \updelta_{D}(k) \left(\frac{1}{N}\right)
\end{equation}
Taking the IFT of the isotropic power spectrum gives us the isotropic 2PCF 
\begin{equation}
\begin{split}
\xi(r) = \sum_{\rm l=1}^{N} \xi_{\rm \sigma_{\rm l}}(r) - \frac{1}{N}.
\end{split}
\end{equation}
Hence, we define the normalised 2PCF, $\tilde{\xi}(r)$, as
\begin{equation}
\tilde{\xi}(r) = \xi(r) + \frac{1}{N}.
\end{equation}
with an offset of $1/N$ which in the case of a single Gaussian clump reduces to unity. Since the offset is a result of finite number of uncorrelated clumps we have to fit this term to the LS-estimator, which is defined for a mean zero field, to measure the correct value of $r_{\rm peak}$. 

\end{appendix}


\end{document}